\documentclass{aastex631}

\usepackage{amsmath}
\usepackage{mathrsfs}
\usepackage{threeparttable}
\usepackage{float}

\shorttitle{BASSET}
\shortauthors{Cao et al.}
\graphicspath{{./}{figures/}}
\begin{document}

\title{BASSET: Bandpass-Adaptive Single-pulse SEarch Toolkit -- Optimized Sub-Band Pulse Search Strategies for Faint Narrow-Band FRBs}

\author[0009-0000-7501-2215]{J.~H Cao}
\affiliation{National Astronomical Observatories, Chinese Academy of Sciences \\
20A Datun Road, Chaoyang District, Beijing 100101, China}
\affiliation{University of Chinese Academy of Sciences \\
Beijing 100049, China}

\author[0000-0002-3386-7159]{P. Wang}
\affiliation{National Astronomical Observatories, Chinese Academy of Sciences \\
20A Datun Road, Chaoyang District, Beijing 100101, China}
\affiliation{Institute for Frontiers in Astronomy and Astrophysics, Beijing Normal University \\
Beijing 102206, China}
\correspondingauthor{P. Wang}
\email{E-mail: wangpei@nao.cas.cn}

\author[0000-0003-3010-7661]{D. Li}
\affiliation{Department of Astronomy, Tsinghua University \\
Beijing 100084, China}
\affiliation{National Astronomical Observatories, Chinese Academy of Sciences \\
20A Datun Road, Chaoyang District, Beijing 100101, China}
\affiliation{Zhejiang Lab, Hangzhou, Zhejiang 311121, China}
\correspondingauthor{D. Li}
\email{E-mail: dili@mail.tsinghua.edu.cn}

\author{Q.~H. Pan}
\affiliation{Zhejiang Lab, Hangzhou, Zhejiang 311121, China}
\correspondingauthor{Q.~H. Pan}
\email{E-mail: panqh@zhejianglab.org}

\author{K. Mao}
\affiliation{Zhejiang Lab, Hangzhou, Zhejiang 311121, China}

\author[0000-0001-6651-7799]{C.~H. Niu}
\affiliation{Central China Normal University \\
Wuhan 430079, China}

\author[0000-0002-8744-3546]{Y.~K. Zhang}
\affiliation{National Astronomical Observatories, Chinese Academy of Sciences \\
20A Datun Road, Chaoyang District, Beijing 100101, China}
\affiliation{University of Chinese Academy of Sciences \\
Beijing 100049, China}

\author[0009-0005-5413-7664]{Q.~Y. Qu}
\affiliation{National Astronomical Observatories, Chinese Academy of Sciences \\
20A Datun Road, Chaoyang District, Beijing 100101, China}
\affiliation{University of Chinese Academy of Sciences \\
Beijing 100049, China}

\author[0000-0001-5653-3787]{W.~J. Lu}
\affiliation{National Astronomical Observatories, Chinese Academy of Sciences \\
20A Datun Road, Chaoyang District, Beijing 100101, China}
\affiliation{University of Chinese Academy of Sciences \\
Beijing 100049, China}

\author[0009-0005-8586-3001]{J.~S. Zhang}
\affiliation{National Astronomical Observatories, Chinese Academy of Sciences \\
20A Datun Road, Chaoyang District, Beijing 100101, China}
\affiliation{University of Chinese Academy of Sciences \\
Beijing 100049, China}

\author[0009-0009-8320-1484]{Y.~H. Zhu}
\affiliation{National Astronomical Observatories, Chinese Academy of Sciences \\
20A Datun Road, Chaoyang District, Beijing 100101, China}
\affiliation{University of Chinese Academy of Sciences \\
Beijing 100049, China}

\author[0000-0002-7372-4160]{Y.~D. Wang}
\affiliation{National Astronomical Observatories, Chinese Academy of Sciences \\
20A Datun Road, Chaoyang District, Beijing 100101, China}
\affiliation{University of Chinese Academy of Sciences \\
Beijing 100049, China}

\author{H.~X. Chen}
\affiliation{Zhejiang Lab, Hangzhou, Zhejiang 311121, China}

\author[0000-0001-5738-9625]{X.~L. Chen}
\affiliation{National Astronomical Observatories, Chinese Academy of Sciences \\
20A Datun Road, Chaoyang District, Beijing 100101, China}

\author{E. G\"{u}gercino\u{g}lu}
\affiliation{National Astronomical Observatories, Chinese Academy of Sciences \\
20A Datun Road, Chaoyang District, Beijing 100101, China}

\author[0000-0001-9956-6298]{J.~H. Fang}
\affiliation{Zhejiang Lab, Hangzhou, Zhejiang 311121, China}

\author[0000-0002-0475-7479]{Y. Feng}
\affiliation{Zhejiang Lab, Hangzhou, Zhejiang 311121, China}
\affiliation{Institute for Astronomy, School of Physics, Zhejiang University, \\ Hangzhou 310027, China}

\author[0000-0003-2516-6288]{H. Gao}
\affiliation{Institute for Frontiers in Astronomy and Astrophysics, Beijing Normal University \\
Beijing 102206, China}
\affiliation{Department of Astronomy, Beijing Normal University, \\
Beijing 100875, China}

\author[0000-0001-7199-2906]{Y.~F. Huang}
\affiliation{School of Astronomy and Space Science, Nanjing University, \\
Nanjing 210023, China}
\affiliation{Key Laboratory of Modern Astronomy and Astrophysics (Nanjing University), Ministry of Education, \\
Nanjing 210023, China}

\author[0000-0003-1720-9727]{J. Li}
\affiliation{Department of Astronomy, School of Physical Sciences, University of Science and Technology of China \\
Hefei 230026, China}

\author[0000-0002-9441-2190]{C.~C. Miao}
\affiliation{Zhejiang Lab, Hangzhou, Zhejiang 311121, China}

\author[0000-0002-9390-9672]{C.~W. Tsai}
\affiliation{National Astronomical Observatories, Chinese Academy of Sciences \\
20A Datun Road, Chaoyang District, Beijing 100101, China}
\affiliation{Institute for Frontiers in Astronomy and Astrophysics, Beijing Normal University \\
Beijing 102206, China}
\affiliation{University of Chinese Academy of Sciences \\
Beijing 100049, China}

\author{J.~M. Yao}
\affiliation{Xinjiang Astronomical Observatory, Chinese Academy of Sciences \\
Urumqi 830011, China}

\author{S.~P. You}
\affiliation{Key Laboratory of Information and Computing Science Guizhou Province, Guizhou Normal University, \\
Guiyang 550001, China}

\author[0000-0002-1243-0476]{R.~S. Zhao}
\affiliation{Guizhou Provincial Key Laboratory of Radio Astronomy and Data Processing, Guizhou Normal University \\
Guiyang 550001, China}

\author{Q.~Z. Liu}
\affiliation{Purple Mountain Observatory, Chinese Academy of Sciences \\
Nanjing 210023, China}
\affiliation{College of Physics and Electronic Engineering, Qilu Normal University, \\
Jinan 250200, China}

\author[0000-0001-7746-9462]{S.~M. Weng}
\affiliation{Key Laboratory for Laser Plasmas (MoE), School of Physics and Astronomy, Shanghai Jiao Tong University, \\ 
Shanghai 200240, China}

\author[0000-0002-5799-9869]{S.~H. Yew}
\affiliation{Key Laboratory for Laser Plasmas (MoE), School of Physics and Astronomy, Shanghai Jiao Tong University, \\
Shanghai 200240, China}

\author{J. Zhang}
\affiliation{College of Physics and Electronic Engineering, Qilu Normal University, \\
Jinan 250200, China}

\author[0000-0001-8539-4237]{L. Zhang}
\affiliation{National Astronomical Observatories, Chinese Academy of Sciences \\
20A Datun Road, Chaoyang District, Beijing 100101, China}
\affiliation{Centre for Astrophysics and Supercomputing, Swinburne University of Technology \\
P.O. Box 218, Hawthorn, VIC 3122, Australia}

\author[0000-0002-7420-9988]{D.~K. Zhou}
\affiliation{Zhejiang Lab, Hangzhou, Zhejiang 311121, China}

\author[0000-0001-5105-4058]{W.~W. Zhu}
\affiliation{National Astronomical Observatories, Chinese Academy of Sciences \\
20A Datun Road, Chaoyang District, Beijing 100101, China}
\affiliation{Institute for Frontiers in Astronomy and Astrophysics, Beijing Normal University \\
Beijing 102206, China}



\begin{abstract}

The existing single-pulse search algorithms for fast radio bursts (FRBs) do not adequately consider the frequency bandpass pattern of the pulse, rendering them incomplete for the relatively narrow-spectrum detection of pulses.
We present a new search algorithm for narrow-band pulses to update the existing standard pipeline, Bandpass-Adaptive Single-pulse SEarch Toolkit (BASSET).
The BASSET employs a time-frequency correlation analysis to identify and remove the noise involved by the zero-detection frequency band, thereby enhancing the signal-to-noise ratio (SNR) of the pulses. 
The BASSET algorithm was implemented on the FAST real dataset of FRB 20190520B, resulting in the discovery of additional 79 pulses through reprocessing.
The new detection doubles the number of pulses compared to the previously known 75 pulses, bringing the total number of pulses to 154.
In conjunction with the pulse calibration and the Markov Chain Monte Carlo (MCMC) simulated injection experiments, this work updates the quantified parameter space of the detection rate.
Moreover, a parallel-accelerated version of the BASSET code was provided and evaluated through simulation.
BASSET has the capacity of enhancing the detection sensitivity and the SNR of the narrow-band pulses from the existing pipeline, offering high performance and flexible applicability.
BASSET not only enhances the completeness of the low-energy narrow-band pulse detection in a more robust mode, but also has the potential to further elucidate the FRB luminosity function at a wider energy scale.

\end{abstract}



\section{Introduction} \label{sec:intro}
Fast radio bursts (FRBs) are short-duration coherent radio transients with high energy (for reviews, see \cite{petroff2019fast, cordes2019fast, zhang2020physical, xiao2021physics, zhang2023physics}).
The dispersion measures (DMs) of FRBs far exceed the galactic contributions \citep{thornton2013population}, indicating an extragalactic origin.
The highly variable rotation measures (RMs) suggest that the FRB is surrounded by a dynamic magnetized plasma \citep{michilli2018extreme, xu2022fast, mckinven2023large, anna2023magnetic}.

FRB repeaters are continuously observed \citep{spitler2016repeating, amiri2018chime, niu2022repeating}, although more FRBs are one-off events.
The luminosity function or energy distribution is essential for revealing the nature of the FRB repeater engines.
The complex luminosity functions (e.g., FRB 20121102A \citep{li2021bimodal}, FRB 20201124A \citep{zhang2022fast}, FRB 20190520B \citep{niu2022repeating}, and FRB 20220912A \citep{zhangyk2023fast}) imply the existence of multiple radiation mechanisms.
The observed luminosity functions are significantly affected by biases at the low-energy end.
A more intrinsic luminosity function will be obtained by increasing the detection completeness for low-energy pulses.
This is related to the FRB search pipelines.

The standard search pipelines are diverse \citep{petroff2019fast}.
The Parkes and the UTMOST telescopes use the \texttt{HEIMDALL}\footnote{\url{https://sourceforge.net/projects/heimdall-astro/}}.
\texttt{HEIMDALL} employs brute force de-dispersion techniques on GPUs \citep{champion2016five, caleb2017first}.
The Arecibo and Green Bank surveys have applied the \texttt{PRESTO}\footnote{\url{https://www.cv.nrao.edu/~sransom/presto/}} \citep{ransom2001new}, with sub-band de-dispersion techniques \citep{spitler2014fast}.
Pulses were identified in the ASKAP telescope utilizing the \texttt{FREDDA} \citep{bannister2017detection}.
This pipeline is based on the Fast Dispersion Measure Transform (FDMT) \citep{zackay2017accurate}.
Their single-pulse search algorithms are logically similar, generally following the same steps \citep{petroff2019fast}:
\begin{itemize}
    \item Preliminary radio frequency interference (RFI) excision: A portion of RFI can be removed before search. Time samples and frequency channels affected by RFI will be masked.
    \item De-dispersion: Corrects for the dispersion effects caused by cold plasma.
    \item Extracting a time series: The data are averaged over all frequency band to produce one-dimensional time series.
    \item Baseline estimation or smoothing: Removes the non-uniform baseline in the time series, which is caused by variations in the mean signal.
    \item Normalization: Estimates the noise properties to calculate the signal-to-noise ratio (SNR) of a pulse.
    \item Matched filtering: Convolves the time series with box-car functions of various lengths to identify pulses wider than a single time sample. Peaks in the de-dispersed, normalized, and convolved time series are reported as candidates.
    \item Candidate grouping: Clusters single-pulse candidates that are likely related to the same event.
    \item Post-processing RFI excision: Further RFI excision is performed using the candidate list.
\end{itemize}

These pipelines have different detection sensitivities for technical reasons.
For example, \citet{gourdji2019sample} detected 41 pulses, while \citet{aggarwal2021comprehensive} detected 134 pulses by using the same observational data for FRB 20121102A.

The Canadian Hydrogen Intensity Mapping Experiment (CHIME) has reported that repeating FRBs have larger pulse widths and narrower bandwidths compared to the one-off events \citep{pleunis2021fast}.
Pulses from repeating FRBs tend to occupy only a part of the observation bandwidth.
The zero-detection frequency band is also averaged in the \texttt{Extracting a time series} step, involving noise and decreasing the SNR of the pulses.
A new search algorithm is needed to improve the detection completeness for faint narrow-band pulses.

Sub-band search can partially solve this issue, but it depends on the choice of the sub-bands.
\citet{spitler2012multimoment} used a multi-moment technique to quantify the distribution of the pulse intensity across frequencies.
This technique is only used for excluding the narrow-band RFI.
\citet{kumar2024detecting} uses a Kalman detector technique to increase the SNR of candidates.
The Kalman detector assumes that the spectrum follows a Gaussian process, which is hard to account for the morphologically complex FRB pulses.
\citet{men2024transientx} established \texttt{TRANSIENTX}, which utilizes the density-based spatial clustering of applications with noise (DBSCAN) algorithm to remove the duplicate candidates.
The program is CPU-based, posing an issue for GPU users.
Additionally, DBSCAN lacks of being interpretable.

In this paper, we present a new user-friendly search toolkit for \texttt{PRESTO}, the "Bandpass-Adaptive Single-pulse SEarch Toolkit" (BASSET\footnote{\url{https://github.com/caojhNAOC/BASSET}}).
BASSET significantly improves the detection sensitivity for the narrow-band pulses and has the potential to facilitate the search in broad-band observations.
In Section 2, we present the framework and algorithms of BASSET.
In Section 3, we perform a test on BASSET by applying it to observational data made by the Five-hundred-meter Aperture Spherical radio Telescope (FAST) (\citet{nan2011five}, \citet{li2018fast}).
In Section 4, we reprocess the FRB 20190520B data collected by FAST using BASSET.
In Section 5, we update the results from the previous detection by \citet{niu2022repeating}, demonstrate a parallel-accelerated version of BASSET, and conduct MCMC simulation experiments.
Finally, we provide a conclusion in Section 6.

\section{The Bandpass-Adaptive Single-pulse SEarch Toolkit (BASSET)}
\begin{figure}
    \plotone{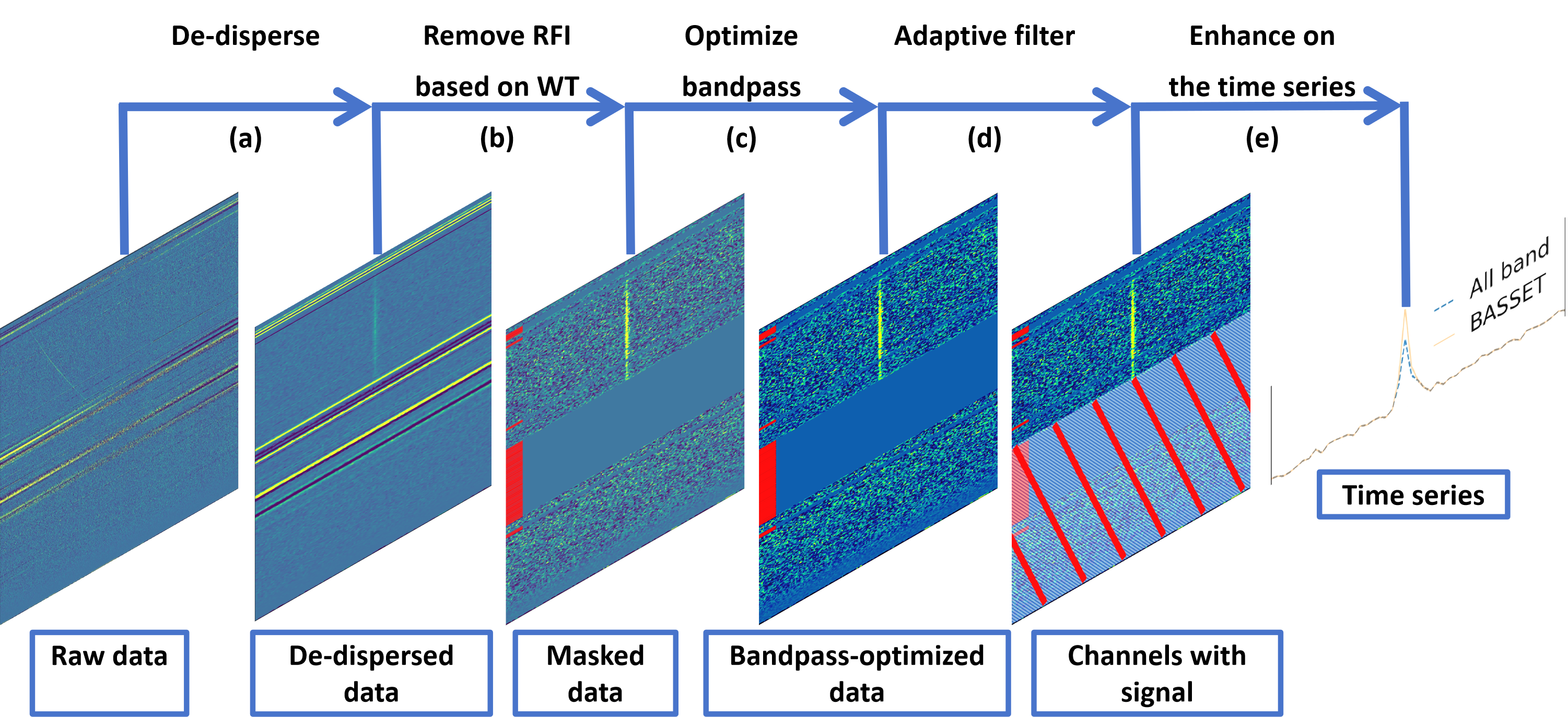}
    \caption{
    The framework of BASSET.
    The SNR of the narrow-band pulses is improved on the time series using BASSET.
    The frequency channels affected by RFI are masked and highlighted.
    }
    \label{fig:framework}
\end{figure}

BASSET is a user-friendly update toolkit for \texttt{PRESTO}.
BASSET is designed to enhance \texttt{PRESTO}'s detection sensitivity for the narrow-band faint pulses.
The \texttt{PRESTO}'s framework with BASSET is shown in Fig.~\ref{fig:framework}, with the following two steps updated:
\begin{itemize}
    \item \texttt{Preliminary RFI excision} by using the wavelet transform (WT) techniques (Fig.~\ref{fig:framework} (b)).
    \item \texttt{Extracting a time series} without noise involved by the zero-detection frequency band.
    The system bandpass spectrum is removed firstly (Fig.~\ref{fig:framework} (c)) as data preparation to load adaptive filter (Fig.~\ref{fig:framework} (d)).
\end{itemize}

\subsection{RFI removal based on wavelet transform}
We masked RFI-contaminated data to reduce the number of false-positive candidates.
The \texttt{rfifind} provided by \texttt{PRESTO} may fail to remove the non-stationary RFI.
We then developed a new RFI removal program based on WT, \texttt{rfi-mask.py}.
This program performs a two-dimensional wavelet decomposition on the dynamic spectrum, removes the components with outlier wavelet coefficients, and generates a mask file like \texttt{rfifind}.
\texttt{rfi-mask.py} has also been added to the BASSET library.

\subsection{Bandpass optimization}
The pulse spectrum in the masked data (Fig.~\ref{fig:framework}) lies on the top of the system bandpass spectrum.
The system bandpass spectrum is caused by the sky, ground, and receiver temperatures, and is related to the filters in the receiver chain \citep{o2002single}.
The system bandpass spectrum has been removed as data preparation to load adaptive filter.
This update has been integrated into the \texttt{prepsubband} of \texttt{PRESTO}.

\subsection{Adaptive filter}
We then perform an adaptive filter on the de-dispersed data after masking RFI and optimizing the bandpass.
The adaptive filter removes the noise involved by the zero-detection frequency band, leading to an increased significance of the pulse detection.
This update has also been integrated into the \texttt{prepsubband}.
The adaptive filter consists of three steps: \texttt{Triggering}, \texttt{Searching Time-Frequency Range}, and \texttt{Enhancement}.
The pulse candidates and their time of arrival (ToA) are obtained based on the bandpass information in the \texttt{Triggering} step.
Their time-frequency ranges are then localized in the \texttt{Searching Time-Frequency Range} step.
Subsequently, the SNR of the candidates is improved in the \texttt{Enhancement} step.
The algorithm for the three steps is as follows:

\begin{itemize}
    \item Triggering:
    
    The spectra of the narrow-band FRB pulses are best described by a box-car or a Gaussian function \citep{pleunis2021fast}.
    We describe the ideal de-dispersed pulse spectral intensity $I(f)$ as:
    \begin{equation}
    \label{eq:model}
    I(f) = 
    \begin{cases} 
        I_{0}, & \text{if } CF - BW / 2 < f < CF + BW / 2, \\
        0, & \text{otherwise}, 
    \end{cases}
    \end{equation}
    where $f$ is the channel index ranging from $0$ to $N - 1$, and $N$ is the total number of the frequency channels. $I_0$ is the characteristic intensity of the pulse. $BW$ is the bandwidth of the pulse, and $CF$ is the central frequency of the pulse.
    
    Based on Eq.~\ref{eq:model}, the adaptive filter performs matched filtering on the data's bandpass every 5 ms to get the candidates.
    The matched filtering function $\mathscr{F}(i)$ is a series of box-car functions:
    \begin{equation}
    \mathscr{F}(i) = 
    \begin{cases} 
        1, & \text{if } 0 < i < L, L = L_0, L_1 ...... L_{m}, \\
        0, & \text{otherwise}, 
    \end{cases}
    \end{equation}
    where $L$ is the length of the box-car function, and $L_0, L_1, \ldots, L_m$ are a series of trail lengths used to match the pulses with different bandwidths.
    
    The adaptive filter then uses the autocorrelation function (ACF) of the candidates' bandpass to eliminate some false-positive items:
    \begin{equation}
        ACF(i)= \sum_{k=1}^{N - \left|N-i\right|} I_{Cand.}(k) * I_{Cand.}(k + \left|N-i\right|),
    \end{equation}
    where $i$ ranges from 0 to $2N - 2$, and $Cand.$ is the candidate.
    The $ACF$ is symmetric about $ACF(n)$.
    A Gaussian function ($G(x)$) is used to fit the width of the $ACF$, $W_{ACF}$:
    \begin{equation*}
        G(x) = A e^{-\frac{(x - \mu)^2}{2\sigma^2}},
    \end{equation*}
    with
    \begin{equation}
        W_{ACF} = 2 \sigma.
    \end{equation}
    Candidates with $W_{ACF}$ less than a preset threshold are removed, as $W_{ACF}$ can serve as an estimate of the candidates' bandwidth.
    
    Eq.~(\ref{eq:model}) is an oversimplified description for the morphologically complex FRB pulses.
    From a technical point of view, the matched filtering based on the box-car function is computationally effective.
    The ACF method is model-independent.
    In astronomy, the ACF has been used to measure pulse drift rates by \citet{marthi2022burst}.
    
    \item Searching for the time-frequency range:
    
    The candidates and their ToAs have been obtained in the \texttt{Triggering} step.
    The $CF$ of each candidate can be estimated as:
    \begin{equation}
        CF = \text{max index}(I_{Cand.} \ast \mathscr{F}), L = W_{ACF}.
    \end{equation}
    
    To locate the time range of a candidate, the adaptive filter checks the $W_{ACF}$ and $CF$ for 1 ms time interval adjacent to the candidate.
    When
    \begin{equation}
        \label{eq:time_range}
        \left| CF_{Cand.} - CF_{Adj.} \right| < \frac{\left| W_{ACF \ Cand.} - W_{ACF \ Adj.} \right|}{2},
    \end{equation}
    this time interval is considered to belong to the time range of the candidate.
    Here we use $Cand.$ and $Adj.$ to distinguish the two $ACF$s and $CF$s.
    The next interval is then checked until Eq.~(\ref{eq:time_range}) is not satisfied.

    For a given frequency range, the SNR of the candidate can be expressed as
    \begin{equation}
    \label{eq:frequency_range}
        SNR_{Cand.} = \text{max}(TS_{Cand.}) / \text{STD}(TS_{Bg.}),
    \end{equation}
    where $TS$ is the time series by averaging the data over the given frequency range and $Bg.$ is the background noise data near the candidate.
    The adaptive filter locates the frequency range giving the maximum SNR, $SNR_{Cand.\text{ max}}$, based on Eq.~(\ref{eq:frequency_range}).

    \item Enhancement:
    
    The data within the candidate's time range will be averaged over the localized frequency, increased by a ratio $\beta$:
    \begin{equation}
        \beta = SNR_{Cand.\text{ max}} / SNR_{Cand.\text{ A}},
    \end{equation}
    where $SNR_{Cand.\text{ A}}$ is the SNR of the candidate averaged over all band.
\end{itemize}

\section{Testing BASSET with the FAST data}
\subsection{BASSET performance on single-component pulses}
\begin{figure}
    \plotone{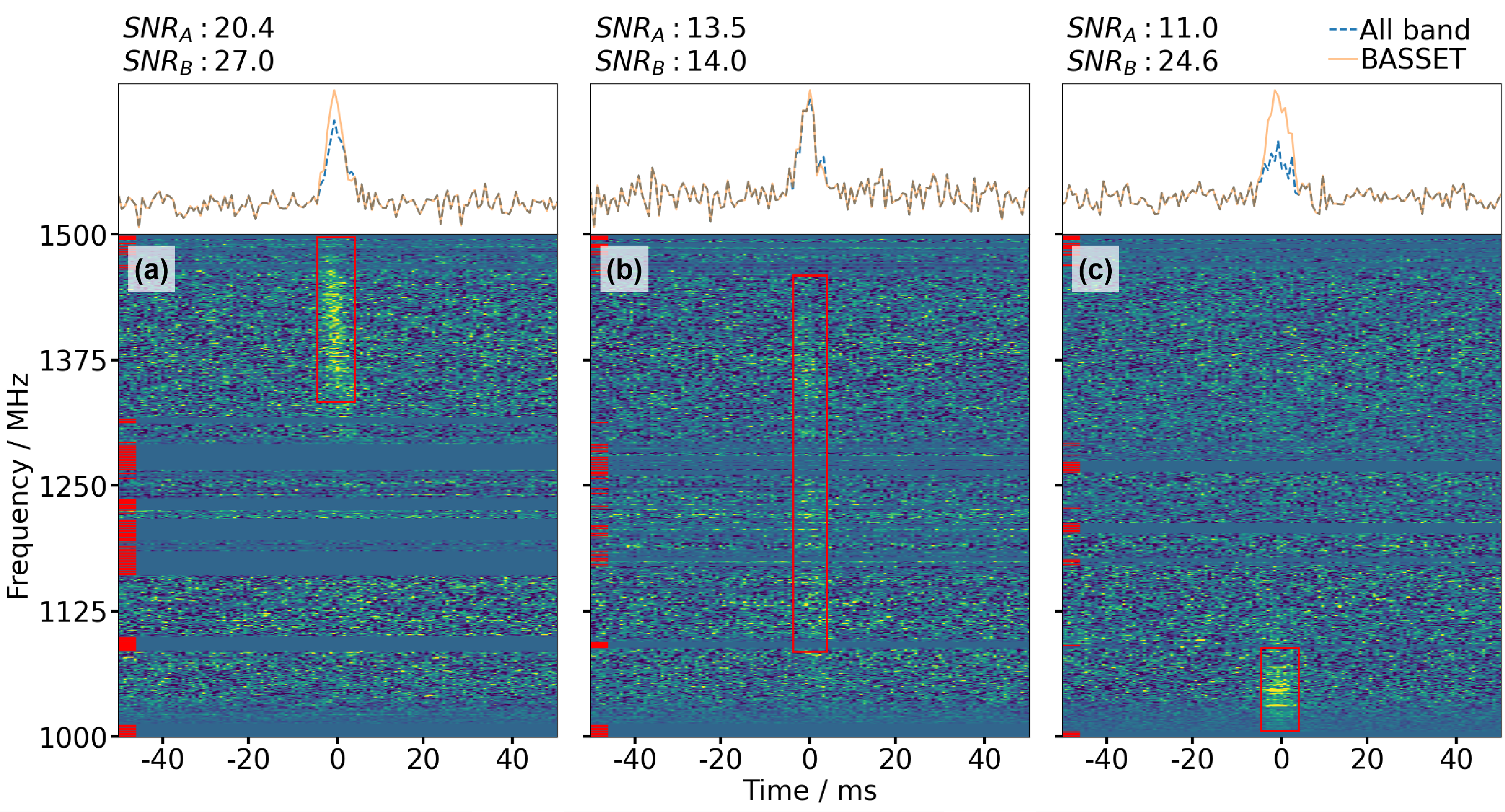}
    \caption{The performance of BASSET from application to the three FAST-detected single-component pulses. The pulses in panels a and b are from FRB 20121102A \citep{li2021bimodal}, while the pulse in panel c is from FRB 20201124A.
    The time-frequency regions searched by BASSET are highlighted with red boxes.
    The frequency channels affected by RFI are masked and highlighted.
    The blue dashed line represents the time series processed by the standard pipeline, and the orange solid line represents the time series processed by BASSET.
    $SNR_A$ and $SNR_B$ represent the maximum SNR obtained by the standard pipeline and BASSET, respectively.
    }
    \label{fig:combined_images}
\end{figure}

We selected a sample set of 627 single-component pulses from FRB 20121102A \citep{li2021bimodal} and FRB 20201124A \citep{zhou2022fast, zhang2022fast, niu2022fast, jiang2022fast} by utilizing the FAST database \texttt{Blinkverse} \cite{xu2023blinkverse}\footnote{\url{https://blinkverse.alkaidos.cn/}}.
We tested BASSET's performance by applying it to this set.
The searched time-frequency ranges are highlighted with red boxes for three of the pulses in Fig.~\ref{fig:combined_images}.
The time series processed by the standard pipeline and BASSET are represented by the blue dashed line and the orange solid line, separately.
The two time series are then convolved with a series of trail box-car functions to obtain the maximum SNR reported as $SNR_\text{A}$ and $SNR_\text{B}$.

BASSET significantly improves the SNR of the narrow-band pulse in Fig.~\ref{fig:combined_images}, panel c, by the ratio $SNR_\text{B}/SNR_\text{A} = 2.24$.
In contrast, this ratio is $SNR_\text{B}/SNR_\text{A} = 1.04$, for the broad-band pulse in Fig.~\ref{fig:combined_images}, panel b.
$SNR_\text{B}/SNR_\text{A}$ is aligned with the fraction of the zero-detection band.

We define the ``optimized time-frequency range" (hereafter referred to as the ``best location").
The pulse yields the maximum SNR ($SNR_{\text{max}}$) when resampled using the best location.
We use "$\text{best}$" to denote the best location and ``$\text{BASSET}$" to denote the time-frequency range obtained by BASSET.
We considered three of typical scenarios based on the bandwidth of the best location ($BW_{\text{best}}$) and the $SNR_{\text{max}}$:
\begin{itemize}
    \item[] \textbf{I.} 389 pulses with broad bandwidths ($BW_{\text{best}}>200\text{ MHz}$)
    \item[] \textbf{II.} 288 pulses with narrow bandwidths ($BW_{\text{best}}<200\text{ MHz}$)
    \item[] \textbf{III.} 29 pulses with narrow bandwidths ($BW_{\text{best}}<200\text{ MHz}$) and low SNR ($SNR_{\text{max}}<15$).
\end{itemize}

\begin{figure}
    \plotone{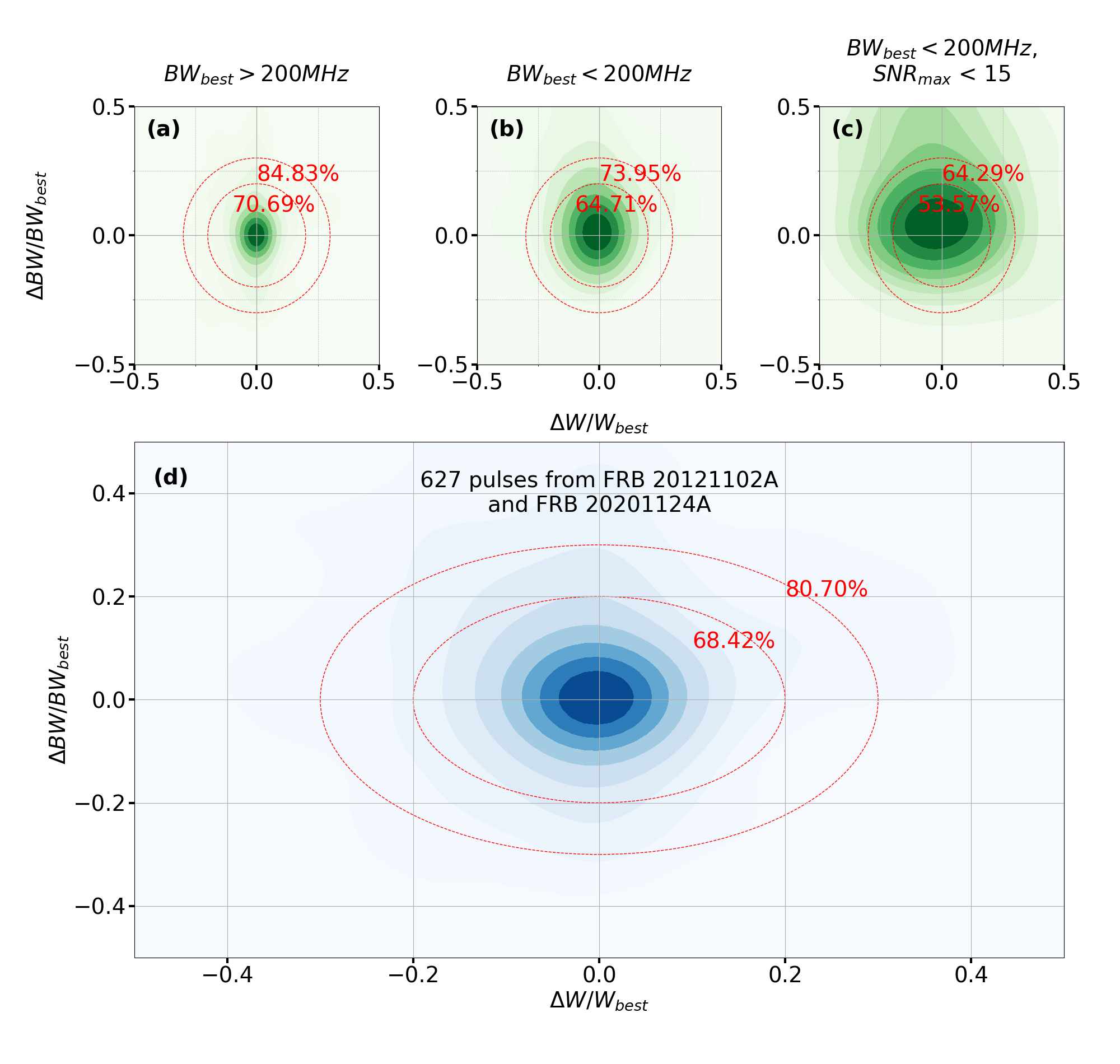}
    \caption{
    The accuracy of \texttt{Searching time-frequency range} is quantified using a 2D KDE of the residuals.
    Panels a, b, and c show the results of the sample subsets with different bandwidths and SNR.
    Panel d shows the result of the entire sample set.
    The percentages of samples with total residuals within 0.2 and 0.3 are shown in red.}
    \label{fig:delta_distribution}
\end{figure}

From a technical point of view, we need to compare the differences between the time-frequency range obtained by BASSET and the best location.
We use the following residual to quantify the accuracy of the \texttt{Searching time-frequency range}:
\begin{equation}
    \Delta BW / BW_{\text{best}} = \frac{BW_{\text{\scriptsize{BASSET}}} - BW_{\text{best}}}{BW_{\text{best}}},
\end{equation}
and
\begin{equation}
    \Delta W / W_{\text{best}} = \frac{W_{\text{\scriptsize{BASSET}}} - W_{\text{best}}}{W_{\text{best}}},
\end{equation}
where $W$ is the width of the time range.
The total residual $r_{\text{total}}$ is:
\begin{equation}
     r_{\text{total}} = \sqrt{{(\Delta BW / BW_{\text{best}}})^2 + ({\Delta W / W_{\text{best}}})^2}.
\end{equation}
We use $SNR_{\text{max}}$ to estimate the uncertainty ($\epsilon$) of the best location:
\begin{equation}
   \epsilon = \frac{1}{\sqrt{SNR_{\text{max}}}}.
\end{equation}
The residual will be scaled by the $\lambda$ factor given by
\begin{equation}
    \lambda = 1 - \frac{\epsilon}{r_{\text{total}}}.
\end{equation}
The residual can be ignored when $\lambda < 0$, as it is within the uncertainty range.
In Fig.~\ref{fig:delta_distribution}, we visualized the residuals of 627 pulses by using the two-dimensional kernel density estimation (2D KDE).
According to the radiometer equation, we can estimate:
\begin{equation}
    \begin{cases} 
    \lambda r_{\text{total}} < 0.2, \, SNR_\text{B} \geq 90\% \, SNR_\text{max} & \text{for 68.42\% completeness}  \\
    \lambda r_{\text{total}} < 0.3, \, SNR_\text{B} \geq 84\% \, SNR_\text{max} & \text{for 80.70\% completeness}
    \end{cases}
\end{equation}
Most of the outlier residual structure in Fig.~\ref{fig:delta_distribution}, panel d is attributed to the frequency drifting or scintillation, leading to blurring the pulse morphology.

\subsection{BASSET performance on complex-structured pulses}
\begin{figure}
    \plotone{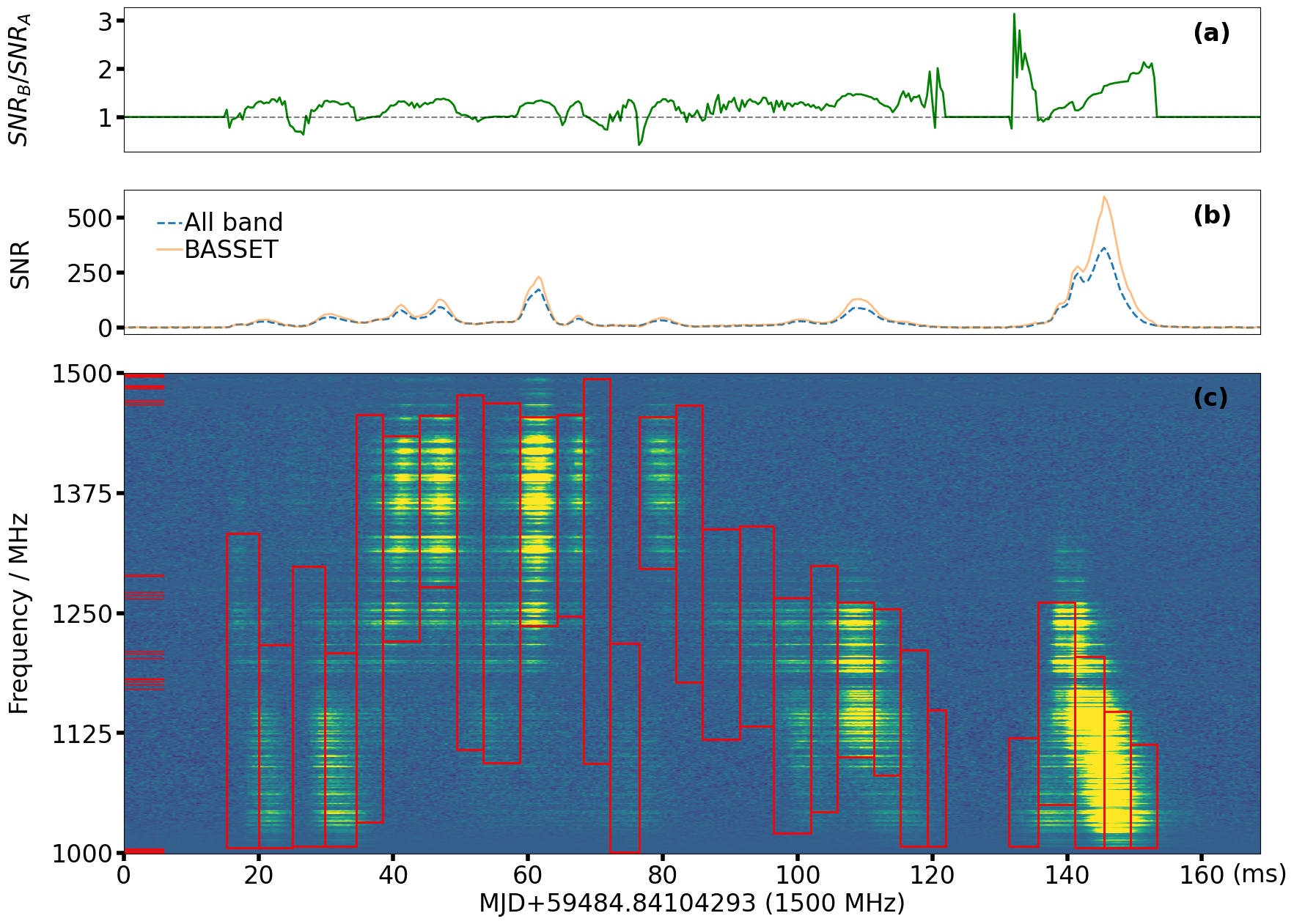}
    \caption{The performance of BASSET when applied to a series of morphologically complex pulses from FRB 20201124A.
    Panel a shows the $SNR_\text{B}/SNR_\text{A}$.
    In panel b, The blue dashed line represents the time series processed by the standard pipeline, and the orange solid line represents the time series processed by BASSET.
    Panel c shows the dynamic spectrum of the pulses.
    The time-frequency regions searched by BASSET are highlighted with red boxes.
    The frequency channels affected by RFI are masked and highlighted.
    }
    \label{fig:FRB20201124A}
\end{figure}

We selected a series of morphologically complex pulses from FRB 20201124A to test the performance of BASSET.
The time-frequency ranges searched by BASSET are shown in Fig.~\ref{fig:FRB20201124A}.
BASSET demonstrates a series of promising candidates, covering all of the pulses (Fig.~\ref{fig:FRB20201124A}, panel c).
The detection significance of the pulses is enhanced, with $SNR_\text{B}/SNR_\text{A} > 1$ (Fig.~\ref{fig:FRB20201124A}, panel a).
This demonstrates the flexibility of BASSET when applying to the morphologically complex pulses.

\subsection{BASSET performance in RFI-contaminated environment}
\begin{figure}
    \plotone{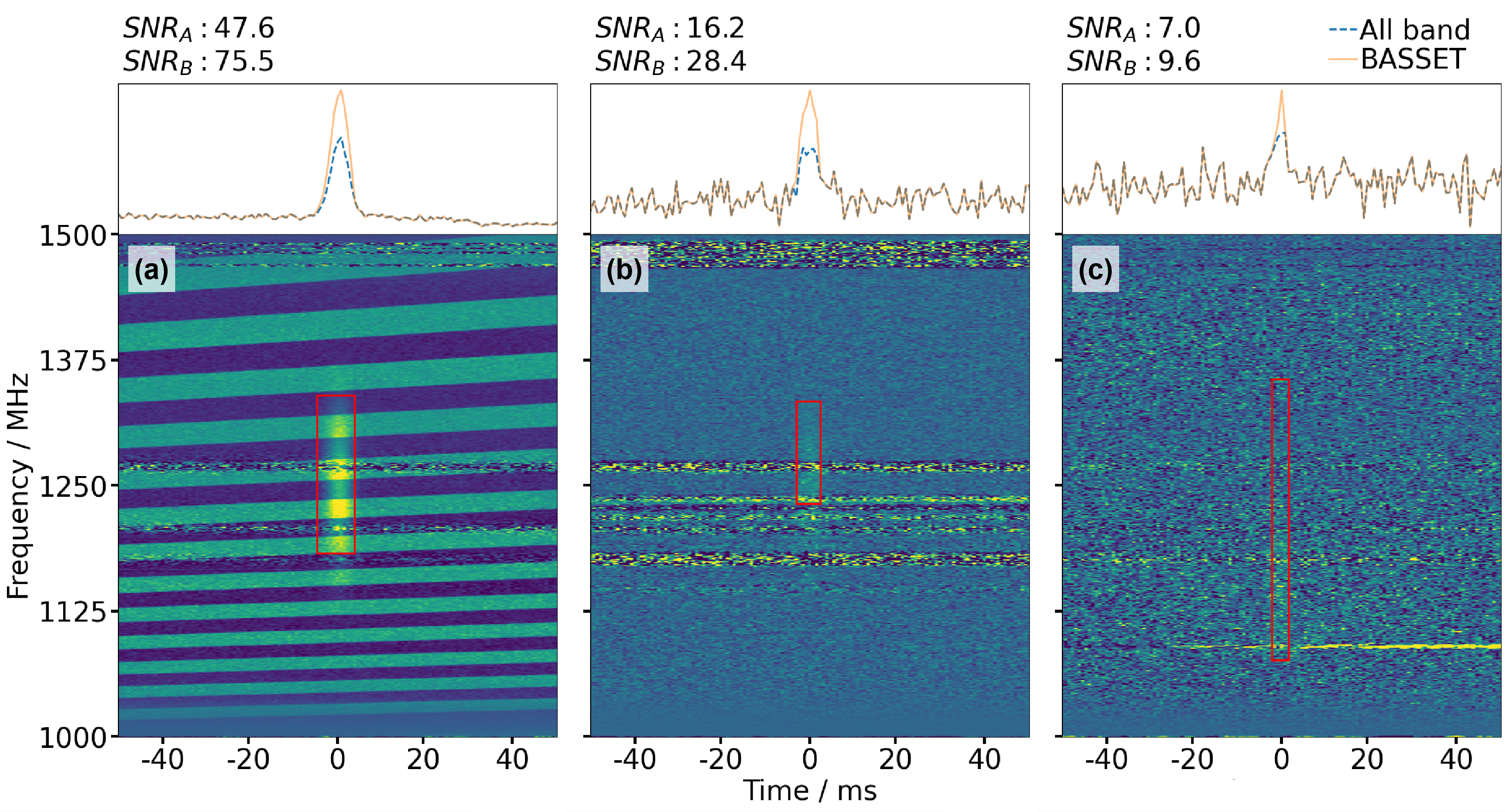}
    \caption{
    BASSET performance in three typical RFI mode environments.
    Panel a: Calibration noise diode. The injected mock pulse has a width of 5 ms and a bandwidth of 300 MHz centered at 1250 MHz, both with a Gaussian shape, and an fluence SNR of 60.
    Panel b: The stationary narrow-band RFI. The injected mock pulse has the same width and bandwidth as panel a, but with an SNR of 10.
    Panel c: The narrow-band RFI like a faint bulge. The injected mock pulse is the same as in panel b with the central frequency changed to 1150 MHz.
    The time-frequency regions searched by BASSET are highlighted with red boxes.}
    \label{fig:rfi}
\end{figure}

In order to test the robustness of BASSET, we designed an experiment by injecting mock pulses into the FAST real data with complicated noise backgrounds.
Three typical RFI modes in FAST data are considered (Fig.~\ref{fig:rfi}):
\begin{itemize}
    \item[] \textbf{I.} Calibration noise diode across all band (Panel a).
    The injected pulse has a width of 5 ms, a bandwidth of 300 MHz centered at 1250 MHz, both with a Gaussian shape, and a fluence SNR of 60 (across all band without removing RFI).
    BASSET will miss pulses with lower SNR in the \texttt{Triggering} step, as the spectral intensity is affected by the noise.

    \item[] \textbf{II.} The stationary narrow-band RFI (Panel b). The injected pulse has the same width and bandwidth as Panel a, but with an SNR of 10.
    
    \item[] \textbf{III.} The narrow-band RFI like a faint bulge (Panel c). The injected pulse is the same as in panel b, except that the central frequency is changed to 1150 MHz.
\end{itemize}

BASSET successfully locates the pulse range and enhances the significance of detection in the three scenarios.
The \texttt{Triggering} step will fail for faint pulses in mode I, indicating that this step is sensitive to the spectral intensity of the pulses.
The $SNR_\text{B}$ decreases to 79.13\%, 58.57\%, and 71.85\% for modes I-III, respectively, compared to the $SNR_\text{B}$ obtained in a clean noise background.
This percentage is aligned with the fraction of the band contaminated by RFI.

\section{Reprocessing FAST observation dataset of FRB 20190520B}
\begin{figure}
    \plotone{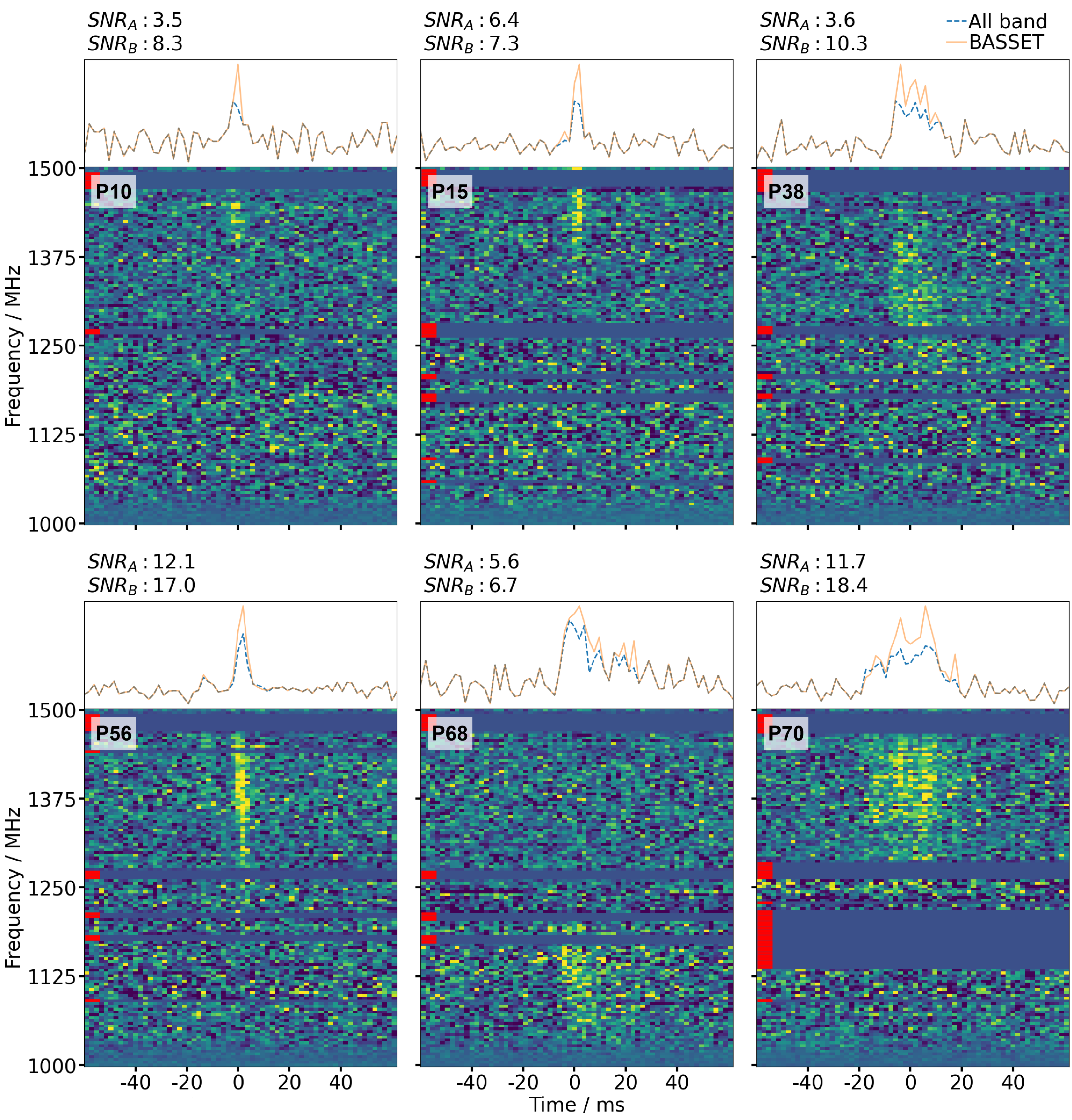}
    \caption{A gallery of the newly detected pulses from FRB 20190520B data reprocessing.
    The $SNR_\text{A}$ is obtained by searching the time series processed by the standard pipeline (blue dashed line), using \texttt{single\_pulse\_search.py} from \texttt{PRESTO}.
    $SNR_\text{B}$ is obtained by searching the time series processed by BASSET (orange solid line).    
    The frequency channels affected by RFI are masked and highlighted.
    }
    \label{fig:new_detected_FRB}
\end{figure}

FAST observations of FRB 20190520B between April and September 2020 were processed using a \texttt{Heimdall}-based pipeline \citep{niu2022repeating}.
Candidates with $SNR_\text{A} > 7$ were recorded, and 75 pulses were detected in 18.5 hours.
There may be missing pulses due to the high detection incompleteness of \texttt{Heimdall} for narrow-band faint pulses.
We reprocessed the data using BASSET.
Candidates with $SNR_\text{B} > 5$ were recorded, and pulses were manually selected using the dynamic spectrum.
The new detection doubles the number of pulses, bringing the total number of pulses to 154.
Six of the newly detected pulses are displayed in Fig.~\ref{fig:new_detected_FRB} and the properties of them are shown in Table~\ref{tab:frb_new_detected}.

\begin{table}
\renewcommand{\arraystretch}{1.5}
\setlength{\tabcolsep}{3pt}
\centering
\caption{
The properties of the six newly detected pulses displayed in Fig.~\ref{fig:new_detected_FRB}.
$BW_\text{Obs.}$ is the observation bandwidth.
More detailed pulse properties are shown in Table~\ref{tab:new_FRB}.}
\label{tab:frb_new_detected}
\begin{tabular}{ccccc}
\hline
Pulse & Bandwidth & $BW/BW_\text{Obs.}$ & $SNR_\text{B}$ & $SNR_\text{A}$ \\
\textit{Num} & (\textit{MHz}) & (\%) & & \\
\hline
10 & $130^{+13}_{-13}$ & 26 & $8.3$ & $3.5$ \\
15 & $131^{+13}_{-13}$   & 26 & $7.3$ & $6.4$ \\
38 & $147^{+15}_{-15}$   & 29 & $10.3$ & $3.6$ \\
56 & $365^{+67}_{-67}$   & 73 & $17.0$ & $12.1$ \\
68 & $194^{+19}_{-19}$   & 39 & $6.7$ & $5.6$ \\
70 & $121^{+12}_{-12}$   & 24 & $18.4$ & $11.7$ \\
\hline
\end{tabular}
\end{table}

Fifty of pulses with $SNR_\text{A} > 5$ missed in the previous search are found by us this time, possibly because we chose a lower SNR threshold.
The detection of 29 pulses with $SNR_\text{A} < 5$ clearly demonstrates the improved detection sensitivity using BASSET.

\begin{figure}
    \plotone{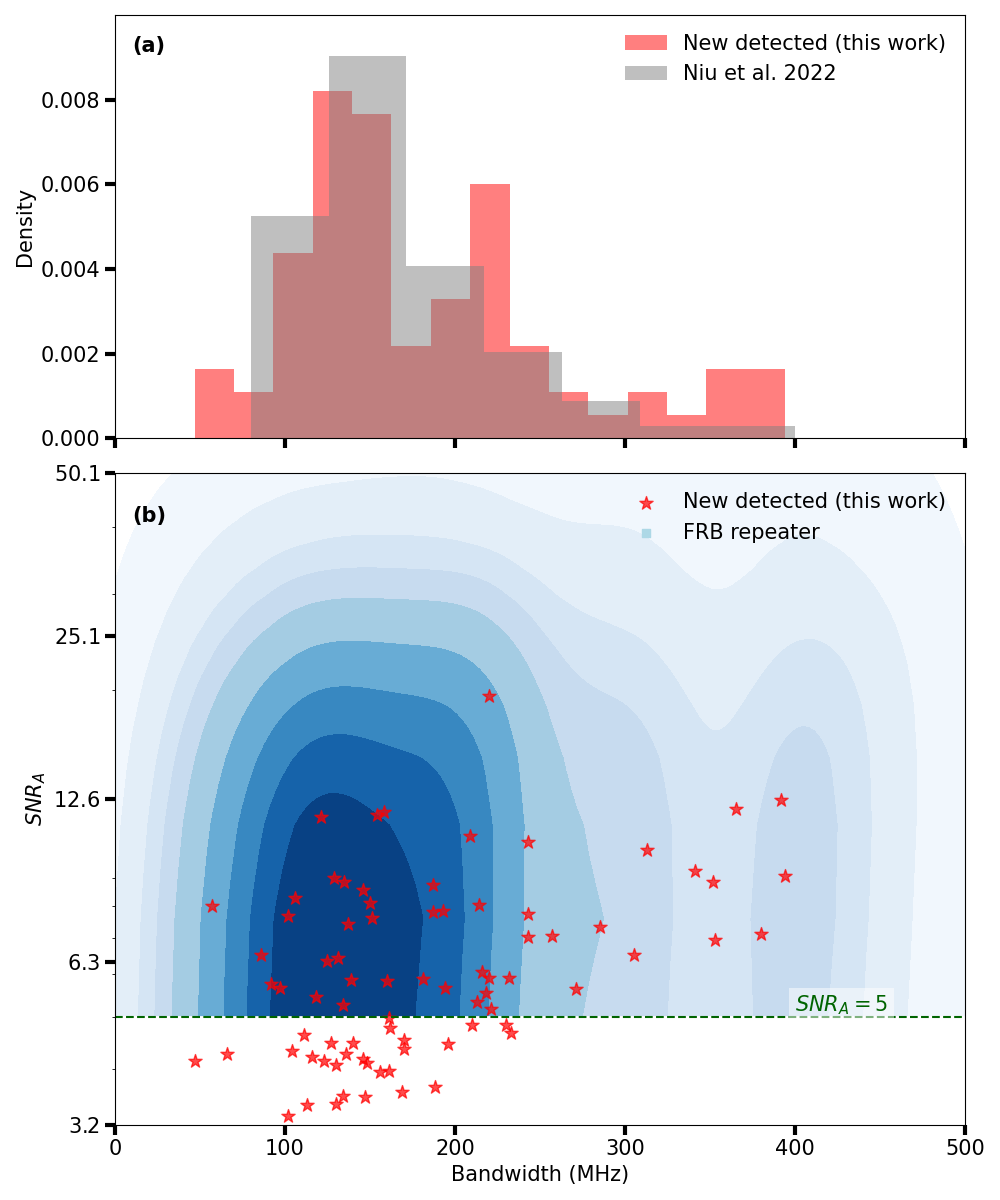}
    \caption{
    The bandwidth and SNR distribution of the newly detected pulses.
    Panel a shows the bandwidth histogram of the newly detected pulses by BASSET, compared to the previous detection by \citet{niu2022repeating}.
    Panel b shows the bandwidth-SNR statistics.
    The newly detected pulses are marked with red stars, compared to the 2D KDE distribution of FRB 20121102A \citep{li2021bimodal}, utilizing the FAST database \texttt{Blinkverse} \citep{xu2023blinkverse}.
    Other FRB repeaters are not considered due to the lack of SNR or bandwidth data.} 
    \label{fig:distribution}
\end{figure}

The bandwidth and SNR distribution of the newly detected pulses with a comparison to the previous detection by \citet{niu2022repeating} are shown in Fig.~\ref{fig:distribution}.
The pulses from FRB 20190520B tend to have narrow emission bandwidths (Fig.~\ref{fig:distribution}, panel a).
The standard pipelines are likely to miss narrow-band faint pulses (Fig.~\ref{fig:distribution}, panel b), as suggested by the 49 of newly detected pulses with $SNR_\text{A} < 10$ and $BW < 200 \text{ MHz}$.

\section{Discussion}
\subsection{The energy distribution of FRB 20190520B}
\begin{figure}
    \centering
    \scalebox{0.8}{\plotone{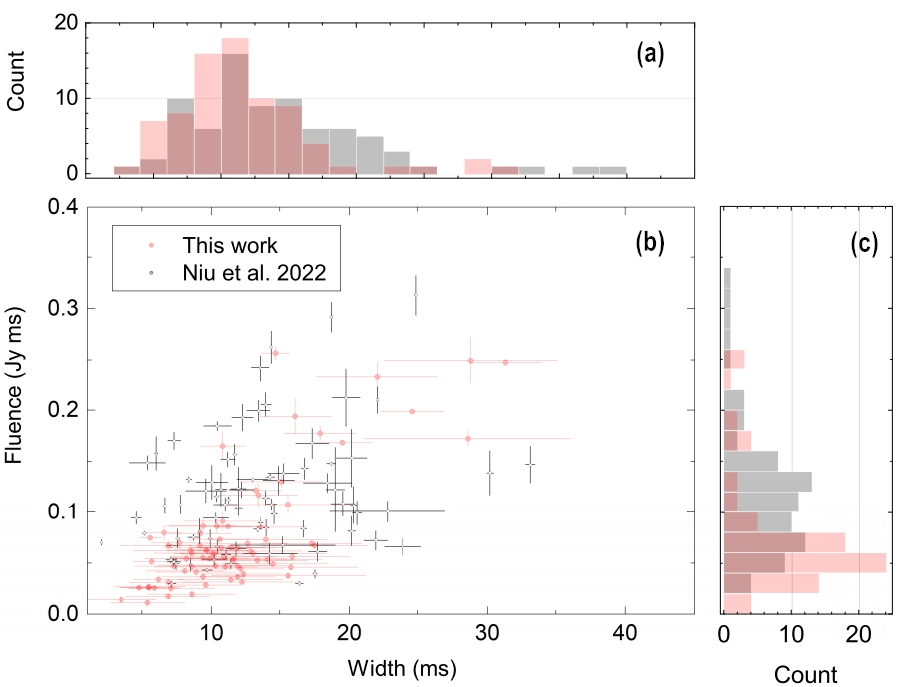}}
    \caption{The fluence-width distribution of the FRB 20190520B. The red dots and histograms are newly detected 79 pulses using BASSET, while the grey dots and bars are taken from \citet{niu2022repeating}.
    } 
    \label{fig:fluence-width}
\end{figure}

\begin{figure}
    \centering
    \scalebox{0.8}{\plotone{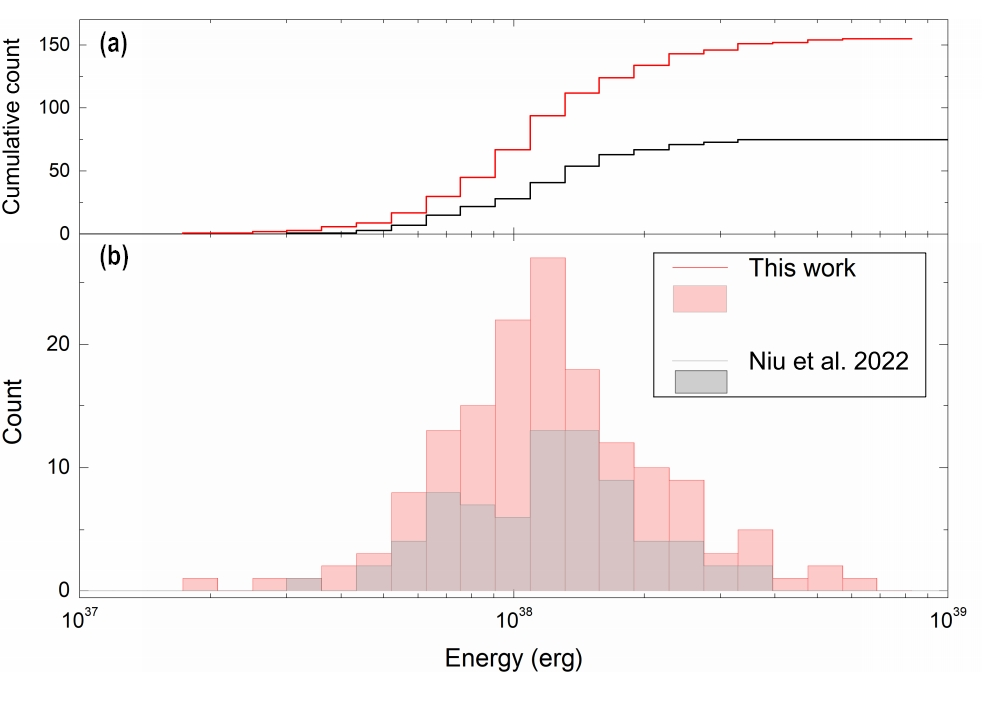}}
    \caption{The luminosity function of the FRB 20190520B. The red line and bars represent the 79 newly detected pulses using BASSET, while the grey line and bars are taken from \citet{niu2022repeating}.
    } 
    \label{fig:energy}
\end{figure}

We updated the energy distribution of the FRB 20190520B reported by \citet{niu2022repeating} using the new detection set.
Following \citet{niu2022repeating}, we used equation (9) of \citet{zhang2018fast}:
\begin{equation}
\label{eq:energy}
    E = 10^{39} \, \text{erg} \times \frac{4\pi}{(1 + z)} \left( \frac{D_L}{10^{28} \, \text{cm}} \right)^2 \left( \frac{F_\nu}{\text{Jy} \cdot \text{ms}} \right) \left( \frac{\nu_c}{\text{GHz}} \right).
\end{equation}
The specific fluence $F_\nu$ is expressed in units of $\text{Jy} \cdot \text{ms}$, measured over the full observation bandwidth. Based on the standard cosmological parameters \citep{macquart2020census}, the luminosity distance for FRB 20190520B is $D_L = 1.218 \, \text{Mpc}$, corresponding to a redshift of $z = 0.241$.
The observation center frequency is $v_c = 1.25 \, \text{GHz}$ for FAST.
The fluence-width distribution is shown in Fig.~\ref{fig:fluence-width} and the luminosity function is presented in Fig.~\ref{fig:energy}.
The luminosity function of FRB 20190520B still exhibits a single modal distribution, suggesting that this may be intrinsic.

\subsection{Parallel-accelerated version}
We implemented a GPU plus OpenMP parallelized version of BASSET to accelerate the FRB search process. We utilize GPU acceleration to parallelize the \texttt{De-dispersion} step and various computations within the \texttt{Triggering} step, including the optimizing calculation, matched filter calculation, and spectrum ACF calculation. For the Gaussian function of the \texttt{Triggering} step and \texttt{Searching time-frequency range} step, OpenMP is applied to accelerate the computation on multicore CPUs.

We compared the performance of our parallelized version with the non-parallelized version of BASSET. The test was performed on a GPU server with 20 Intel i7-12700K cores and one NVIDIA A100-PCIE-40GB GPU, and the software environment was Ubuntu 18.04 with CUDA 11.7. We selected a 140-second segment of real FAST data without pulses, which is single-polarized with a sample time of 98.304 $\mu$s and 4096 frequency channels. The overall performance of the parallelized version showed a 5.98X speedup compared to the non-parallelized version, as shown in Table~\ref{tab:gpu}. The parallel performance of each part in the parallel-accelerated version is shown in Table~\ref{tab:step}.

\begin{table}
\renewcommand{\arraystretch}{1.5}
\setlength{\tabcolsep}{3pt}
\centering
\caption{Comparison of the speed time between the standard pipeline (\texttt{PRESTO}) and the BASSET.$^{a)}$}
\label{tab:gpu}
\begin{tabular}{cc}
\hline
\textbf{Method} & \textbf{Speed Time} \\
& (\textit{s})\\
\hline
PRESTO & 91.28 \\
BASSET(non-parallelized) & 800.45 \\
BASSET(parallelized) & 133.91 \\
\hline
\end{tabular}

\begin{tablenotes}
\item[a)] $a)$ 140-second real FAST data (single-polarized, with a sample time of 98.304 $\mu$s and 4096 frequency channels) without any pulses is used for testing. The test environment consisted of a GPU server with 20 Intel i7-12700K cores and one NVIDIA A100-PCIE-40GB GPU, running on Ubuntu 18.04 with CUDA 11.7.\\%
\end{tablenotes}
\end{table}

\begin{table}
\renewcommand{\arraystretch}{1.5}
\setlength{\tabcolsep}{3pt}
\centering
\caption{The parallel performance of each part in the parallelized version.$^{a)}$}
\label{tab:step}
\small
\begin{tabular}{lccc}
\hline
\textbf{Step} & \textbf{BASSET} & \textbf{BASSET} & \textbf{Improvement} \\
& (\textit{non-parallelized, s}) & (\textit{parallelized, s})\\
\hline
De-dispersion of Data$^{b)}$ & 258.79 & 10.26 & 25.22x \\
Triggering$^{c)}$ & 345.34 & 27.10 & 12.74x \\
\parbox[t]{1.5in}{Searching Time-\\Frequency Range$^{c)}$} & 140.34 & 41.38 & 3.39x \\
\\[-0.4cm]
\hline
\end{tabular}

\begin{tablenotes}
\item[a)] $a)$ 140-second real FAST data (single-polarized, with a sample time of 98.304 $\mu$s and 4096 frequency channels) without any pulses is used for testing. The test environment consisted of a GPU server with 20 Intel i7-12700K cores and one NVIDIA A100-PCIE-40GB GPU, running on Ubuntu 18.04 with CUDA 11.7.\\%
\item[b)] $b)$ Corresponding to "De-disperse" in the framework (Fig.~\ref{fig:framework}).\\%
\item[b)] $c)$ Corresponding to "Adaptive filter" in the framework (Fig.~\ref{fig:framework}).\\%
\end{tablenotes}

\end{table}

\subsection{Comparison between the BASSET and the PRESTO}
In conjunction with pulse calibration and MCMC simulated injection experiments, we update the quantified parameter space of the detection rate for BASSET, compared to the standard pipeline.
The injected mock pulses are Gaussian-shaped in both time and frequency, and are injected into a background of Gaussian noise.
The parameters for the 2000 injected mock pulses are as follows:
\begin{itemize}
    \item \textbf{Fluence SNR} (Uniform distribution):
    \begin{itemize}
        \item Minimum value: 1
        \item Maximum value: 11
    \end{itemize}
    
    \item \textbf{Pulse Width} (Lognormal distribution):
    \begin{itemize}
        \item Mean: 3 ms
        \item STD: 2 ms
    \end{itemize}

    \item \textbf{Bandwidth} (Normal distribution):
    \begin{itemize}
        \item Mean: 150 MHz
        \item STD: 100 MHz
    \end{itemize}

    \item \textbf{Central Frequency} (Normal distribution):
    \begin{itemize}
        \item Mean: 1250 MHz
        \item STD: 100 MHz
    \end{itemize}
    
\end{itemize}
The SNR of the mock pulses is close to the detection limit, as these pulses are likely to be missed during the search.

\begin{figure}
    \plotone{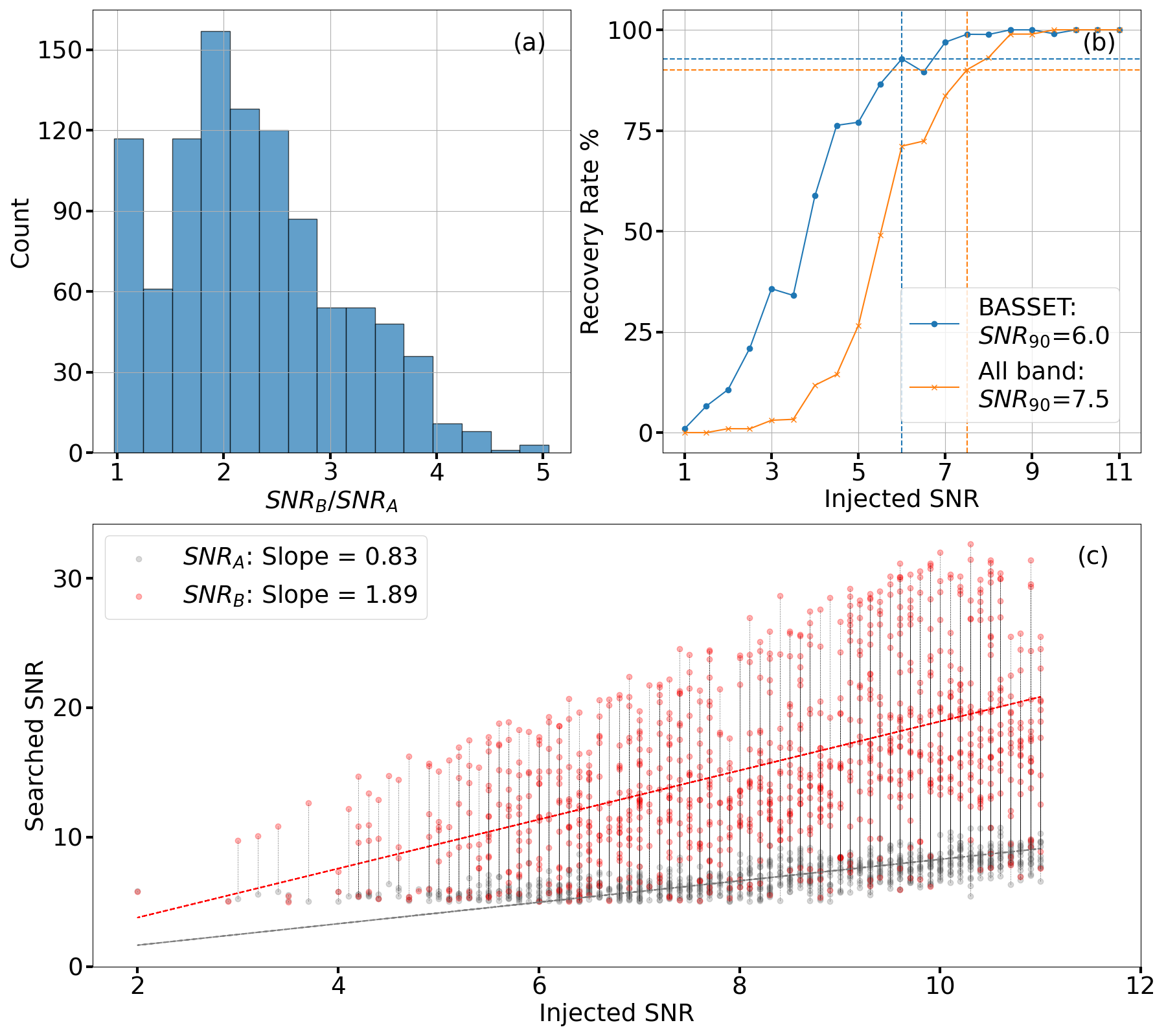}
    \caption{
    Comparison between the BASSET and the standard pipeline.
    The result is based on 2000 mock pulses.
    Panel a presents the distribution of the SNR improvement ratio ($SNR_\text{A}/SNR_\text{B}$).
    Panel b shows the recovery rate as a function of the injected SNR, where $SNR_{90}$ represents the 90\% completeness threshold.
    Panel c shows a comparison between $SNR_\text{A}$ and $SNR_\text{B}$.
    } 
    \label{fig:sim}
\end{figure}

The experimental results are shown in Fig.~\ref{fig:sim}.
BASSET detected 1379 of the mock pulses whereas the standard pipeline detected 1005 of the pulses, with a detection threshold of $SNR=5$.
BASSET enhanced the SNR of 90.91\% of the 1002 pulses detected by both methods (Fig.~\ref{fig:sim}, panel a).
Some faint pulses were missed in the \texttt{Triggering} step, resulting in no enhancement.
The enhancement ratio $SNR_{\text{B}}/SNR_{\text{A}} \approx 2.28$ (Fig.~\ref{fig:sim}, panel c) is consistent with the expected bandwidth (150 MHz).
The 90\% completeness threshold was set from fluence SNR 7.5 to 6.0 using BASSET (Fig.~\ref{fig:sim}, panel b).

\section{Conclusion}
The current single-pulse search algorithms for FRBs are insufficient for the detection of narrow-band pulses due to their failure to account for the frequency bandpass pattern of the pulse.
In this paper, we present a new search algorithm, BASSET, which has been developed to enhance the existing standard pipeline for the detection of narrow-band pulses.
The FAST data for FRB 20190520B were reprocessed using BASSET, and the results of the previous detection by \citet{niu2022repeating} were updated.

\begin{itemize}
    \item We designed BASSET, a user-friendly single-pulse search toolkit for \texttt{PRESTO}, along with a parallel-accelerated version.
    BASSET significantly enhances the detection sensitivity of \texttt{PRESTO} for narrow-band faint pulses.
    \texttt{PRESTO} users can utilize the BASSET by replacing \texttt{prepsubband} with \texttt{prepsubband-BASSET}, and \texttt{rfifind} with \texttt{rfi-mask.py}.

    \item We conducted a series of experiments based on FAST real data reprocessing and MCMC simulations to test the performance of BASSET.
    BASSET enhances the SNR of pulses with different time-frequency morphological structures and demonstrates robustness for complex RFI environments.
    The BASSET search pipeline is more sensitive than the full-band search pipelines, and has established a 90\% completeness threshold, which has lowered the fluence SNR from 7.5 to 6.0.
    
    \item We reprocessed the FAST real dataset of FRB 20190520B using BASSET.
    The newly detected 79 pulses doubled the number of pulses compared to the previously known 75 pulses, bringing the total to 154.
    We calibrated the flux densities of the newly detected pulses and updated the energy distribution of FRB 20190520B reported by \citet{niu2022repeating} using the new detections.
    
    \item BASSET has the potential to expand new parameter space and reveal the intrinsic luminosity function of FRBs.
\end{itemize}


\begin{acknowledgments}
This work is supported by National Natural Science Foundation of China (NSFC) Programs (Grant No. 12375236, 12135009, 12203045, 12103013, 12103069, 11988101, 11725313, 11690024, 12041303, 12473047, U1731238, 12233002, 12041306); by CAS International Partnership Program (No. 114-A11KYSB20160008); by CAS Strategic Priority Research Program (No. XDB23000000); by the National Key R\&D Program of China (No. 2017YFA0402600, 2021YFA0718500); by the National SKA Program of China (No. 2020SKA0120200, 2022SKA0130100, 2020SKA0120300); by the Leading Innovation and Entrepreneurship Team of Zhejiang Province of China (Grant No. 2023R01008); by the Key R\&D Program of Zhejiang (Grant No. 2024SSYS0012) and Foundation of Guizhou Provincial Education Department (Grants No. KY(2023)059).
D.L. is a New Cornerstone investigator.
P.W. acknowledges support from the CAS Youth Interdisciplinary Team, the Youth Innovation Promotion Association CAS (id. 2021055), and the Cultivation Project for FAST Scientific Payoff and Research Achievement of CAMS-CAS.
E.G. is supported by NSFC programme 11988101 under the foreign talents grant QN2023061004L.

This work made use of data from FAST, a Chinese national mega-science facility built and operated by the National Astronomical Observatories, Chinese Academy of Sciences.
\end{acknowledgments}

\section*{Data Availability}
The data is available at \href{https://doi.org/10.57760/sciencedb.15289}{https://doi.org/10.57760/sciencedb.15289}.


\vspace{5mm}
\facilities{FAST:500m}


\software{astropy \citep{2013A&A...558A..33A,2018AJ....156..123A},
PRESTO\citep{ransom2001new},
PyAstronomy\citep{pya},
PyEphem\citep{rhodes2011pyephem}
          }



\appendix

\section{The newly detected pulses from FRB 20190520B}
\begin{figure}[H]
    \plotone{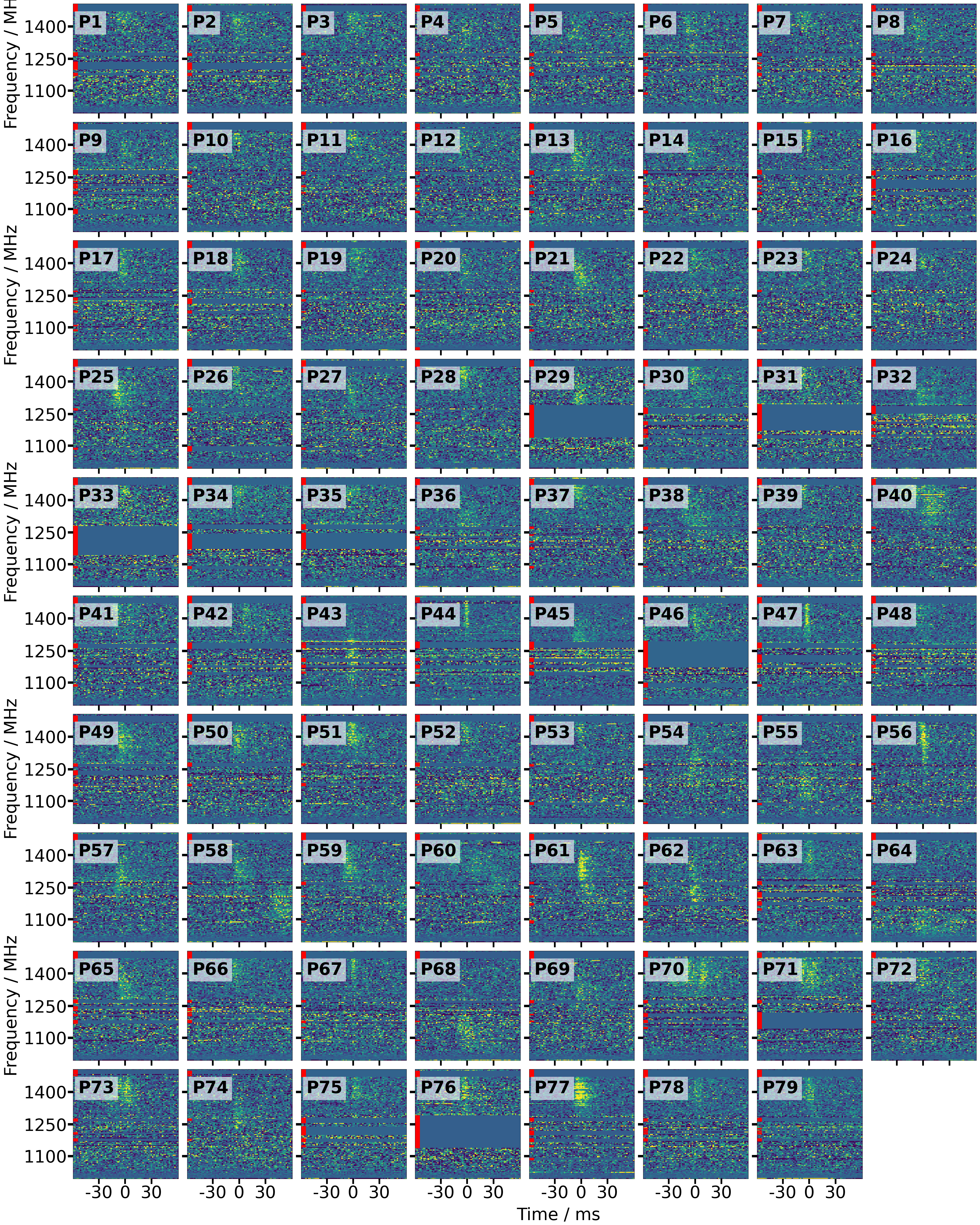}
    \caption{The dynamic spectra of the newly detected pulses from FRB 20190520B.
    The frequency channels affected by RFI are masked and highlighted.}
    \label{fig:new_FRB}
\end{figure}

\renewcommand{\arraystretch}{1.5}
\setlength{\tabcolsep}{3pt}
\begin{center}
\begin{longtable}[H]{c c c c c c c c c c c}
\caption{The properties of the newly detected pulses from FRB 20190520B.}
\label{tab:new_FRB}
\\
\hline
\hline
Pulse & MJD $^{a)}$ & DM$^{b)}$ & W$_{eq}^{c)}$ & Cntrl freq. & Bandwidth$^{d)}$ & Peak flux & Fluence & Energy$_{\Delta \nu}^{e)}$ & SNR$_{B}^{f)}$ & SNR$_{A}^{f)}$ \\
\textit{Num}&(\textit{at infinite freq.}) & (\textit{pc\ cm$^{-3}$}) & (\textit{ms}) & (\textit{MHz}) & (\textit{MHz}) & (\textit{mJy}) & (\textit{mJy\ ms})  & (\textit{$\times 10^{37}$ erg}) \\
\hline
\endhead
\hline
\endfoot
\hline
\endlastfoot
1 & $58963.77015925$ & $1201.84^{+1.98}_{-1.98}$ & $11.9^{+2.8}_{-2.8}$ & $1423$ & $148^{+15}_{-15}$ (\textsuperscript{2}) & $16.19^{+0.52}_{-0.52}$ & $193.14^{+3.94}_{-3.94}$ & $4.19^{+2.64}_{-2.64}$ & $7.1$ & $4.1$ \\%
2 & $58963.77084026$ & $1202.11^{+1.89}_{-1.89}$ & $11.0^{+2.3}_{-2.3}$ & $1436$ & $160^{+16}_{-16}$ (\textsuperscript{1}) & $20.28^{+0.65}_{-0.65}$ & $223.52^{+4.58}_{-4.58}$ & $5.23^{+0.50}_{-0.50}$ & $8.6$ & $5.8$ \\%
3 & $58963.80810873$ & $1202.00^{+2.17}_{-2.17}$ & $15.6^{+5.6}_{-5.6}$ & $1356$ & $134^{+13}_{-13}$ (\textsuperscript{2}) & $9.51^{+0.28}_{-0.28}$ & $148.41^{+2.80}_{-2.80}$ & $2.90^{+2.31}_{-2.31}$ & $6.3$ & $3.6$ \\%
4 & $58991.67612426$ & $1205.57^{+0.28}_{-0.28}$ & $15.8^{+4.8}_{-4.8}$ & $1306$ & $118^{+12}_{-12}$ (\textsuperscript{1}) & $18.09^{+0.69}_{-0.69}$ & $286.26^{+6.92}_{-6.92}$ & $4.95^{+0.87}_{-0.87}$ & $7.1$ & $5.4$ \\%
5 & $58991.67809483$ & $1202.51^{+1.42}_{-1.42}$ & $6.9^{+3.0}_{-3.0}$ & $1368$ & $111^{+11}_{-11}$ (\textsuperscript{1}) & $12.16^{+0.34}_{-0.34}$ & $84.37^{+1.51}_{-1.51}$ & $1.37^{+0.11}_{-0.11}$ & $5.7$ & $4.6$ \\%
6 & $58991.68165178$ & $1205.03^{+1.75}_{-1.75}$ & $12.6^{+3.0}_{-3.0}$ & $1398$ & $220^{+22}_{-22}$ (\textsuperscript{1}) & $17.47^{+1.70}_{-1.70}$ & $220.24^{+13.62}_{-13.62}$ & $7.09^{+0.84}_{-0.84}$ & $7.4$ & $5.9$ \\%
7 & $58991.68701641$ & $1202.32^{+1.58}_{-1.58}$ & $12.1^{+3.2}_{-3.2}$ & $1414$ & $134^{+13}_{-13}$ (\textsuperscript{1}) & $18.35^{+0.59}_{-0.59}$ & $222.87^{+4.54}_{-4.54}$ & $4.38^{+0.50}_{-0.50}$ & $7.5$ & $5.3$ \\%
8 & $58991.68783290$ & $1201.77^{+1.74}_{-1.74}$ & $14.5^{+4.0}_{-4.0}$ & $1410$ & $162^{+16}_{-16}$ (\textsuperscript{2}) & $15.54^{+0.51}_{-0.51}$ & $225.10^{+4.68}_{-4.68}$ & $5.32^{+0.51}_{-0.51}$ & $6.0$ & $4.8$ \\%
9 & $58991.70118536$ & $1202.51^{+1.62}_{-1.62}$ & $5.4^{+2.7}_{-2.7}$ & $1352$ & $66^{+7}_{-7}$ (\textsuperscript{1}) & $17.81^{+0.42}_{-0.42}$ & $96.59^{+1.44}_{-1.44}$ & $0.94^{+0.27}_{-0.27}$ & $6.0$ & $4.3$ \\%
10 & $58991.73307144$ & $1204.62^{+0.82}_{-0.82}$ & $9.6^{+2.5}_{-2.5}$ & $1435$ & $130^{+13}_{-13}$ (\textsuperscript{1}) & $19.02^{+0.64}_{-0.64}$ & $182.74^{+3.90}_{-3.90}$ & $3.48^{+0.31}_{-0.31}$ & $8.3$ & $3.5$ \\%
11 & $58991.73947255$ & $1201.25^{+2.08}_{-2.08}$ & $12.3^{+3.0}_{-3.0}$ & $1428$ & $104^{+10}_{-10}$ (\textsuperscript{1}) & $22.85^{+0.72}_{-0.72}$ & $281.46^{+5.63}_{-5.63}$ & $4.29^{+0.31}_{-0.31}$ & $7.7$ & $4.3$ \\%
12 & $59060.47893855$ & $1203.00^{+1.92}_{-1.92}$ & $10.9^{+3.8}_{-3.8}$ & $1401$ & $169^{+17}_{-17}$ (\textsuperscript{2}) & $19.39^{+0.65}_{-0.65}$ & $212.25^{+9.04}_{-9.04}$ & $5.24^{+0.57}_{-0.57}$ & $6.4$ & $3.6$ \\%
13 & $59060.48025216$ & $1202.73^{+1.65}_{-1.65}$ & $10.6^{+2.2}_{-2.2}$ & $1355$ & $102^{+10}_{-10}$ (\textsuperscript{2}) & $34.92^{+0.91}_{-0.91}$ & $370.64^{+12.24}_{-12.24}$ & $5.53^{+1.54}_{-1.54}$ & $10.6$ & $7.7$ \\%
14 & $59060.48341458$ & $1201.20^{+1.49}_{-1.49}$ & $10.7^{+4.1}_{-4.1}$ & $1350$ & $170^{+17}_{-17}$ (\textsuperscript{2}) & $20.05^{+0.66}_{-0.66}$ & $214.20^{+8.98}_{-8.98}$ & $5.33^{+2.45}_{-2.45}$ & $5.8$ & $4.5$ \\%
15 & $59060.48360265$ & $1200.23^{+1.44}_{-1.44}$ & $4.8^{+1.5}_{-1.5}$ & $1431$ & $131^{+13}_{-13}$ (\textsuperscript{2}) & $32.97^{+1.07}_{-1.07}$ & $159.19^{+6.55}_{-6.55}$ & $3.04^{+0.58}_{-0.58}$ & $7.3$ & $6.4$ \\%
16 & $59060.50838923$ & $1201.65^{+1.88}_{-1.88}$ & $9.9^{+2.8}_{-2.8}$ & $1407$ & $161^{+16}_{-16}$ (\textsuperscript{1}) & $22.90^{+0.77}_{-0.77}$ & $225.65^{+9.69}_{-9.69}$ & $5.32^{+0.57}_{-0.57}$ & $5.6$ & $5.0$ \\%
17 & $59060.52836586$ & $1201.57^{+1.98}_{-1.98}$ & $7.7^{+2.0}_{-2.0}$ & $1385$ & $243^{+24}_{-24}$ (\textsuperscript{1}) & $27.37^{+1.77}_{-1.77}$ & $211.50^{+17.36}_{-17.36}$ & $7.50^{+0.82}_{-0.82}$ & $8.9$ & $7.7$ \\%
18 & $59060.52872694$ & $1201.79^{+1.80}_{-1.80}$ & $10.8^{+2.8}_{-2.8}$ & $1379$ & $285^{+28}_{-28}$ (\textsuperscript{1}) & $23.23^{+1.78}_{-1.78}$ & $250.38^{+24.43}_{-24.43}$ & $10.43^{+1.27}_{-1.27}$ & $9.9$ & $7.3$ \\%
19 & $59060.52955930$ & $1202.43^{+2.07}_{-2.07}$ & $8.1^{+2.8}_{-2.8}$ & $1437$ & $210^{+21}_{-21}$ (\textsuperscript{1}) & $16.65^{+0.61}_{-0.61}$ & $135.26^{+6.29}_{-6.29}$ & $4.15^{+0.50}_{-0.50}$ & $8.5$ & $4.8$ \\%
20 & $59060.53006891$ & $1201.95^{+1.82}_{-1.82}$ & $10.1^{+2.8}_{-2.8}$ & $1361$ & $146^{+15}_{-15}$ (\textsuperscript{2}) & $20.91^{+0.69}_{-0.69}$ & $210.56^{+8.81}_{-8.81}$ & $4.51^{+1.94}_{-1.94}$ & $5.9$ & $4.2$ \\%
21 & $59060.53611424$ & $1202.59^{+1.74}_{-1.74}$ & $13.4^{+2.9}_{-2.9}$ & $1346$ & $313^{+36}_{-36}$ (\textsuperscript{1}) & $23.94^{+1.78}_{-1.78}$ & $321.33^{+30.27}_{-30.27}$ & $14.72^{+1.57}_{-1.57}$ & $14.2$ & $10.1$ \\%
22 & $59060.53668616$ & $1201.06^{+1.48}_{-1.48}$ & $14.0^{+4.4}_{-4.4}$ & $1394$ & $136^{+14}_{-14}$ (\textsuperscript{2}) & $21.55^{+0.74}_{-0.74}$ & $302.59^{+13.24}_{-13.24}$ & $6.01^{+1.55}_{-1.55}$ & $8.5$ & $4.3$ \\%
23 & $59060.54174452$ & $1201.84^{+1.95}_{-1.95}$ & $6.9^{+2.8}_{-2.8}$ & $1408$ & $188^{+19}_{-19}$ (\textsuperscript{1}) & $16.55^{+0.57}_{-0.57}$ & $114.33^{+4.98}_{-4.98}$ & $3.14^{+0.35}_{-0.35}$ & $6.8$ & $3.7$ \\%
24 & $59060.54214026$ & $1202.60^{+1.67}_{-1.67}$ & $5.5^{+2.0}_{-2.0}$ & $1403$ & $102^{+10}_{-10}$ (\textsuperscript{2}) & $24.54^{+0.82}_{-0.82}$ & $133.87^{+5.65}_{-5.65}$ & $2.01^{+0.24}_{-0.24}$ & $5.9$ & $3.3$ \\%
25 & $59060.54245748$ & $1203.56^{+1.71}_{-1.71}$ & $16.1^{+2.6}_{-2.6}$ & $1357$ & $243^{+27}_{-27}$ (\textsuperscript{1}) & $33.57^{+2.35}_{-2.35}$ & $540.72^{+48.14}_{-48.14}$ & $19.24^{+1.96}_{-1.96}$ & $15.1$ & $10.5$ \\%
26 & $59060.54417158$ & $1201.72^{+1.85}_{-1.85}$ & $11.6^{+3.9}_{-3.9}$ & $1415$ & $156^{+16}_{-16}$ (\textsuperscript{2}) & $21.60^{+0.86}_{-0.86}$ & $250.75^{+12.64}_{-12.64}$ & $5.71^{+0.94}_{-0.94}$ & $8.0$ & $4.0$ \\%
27 & $59060.54615672$ & $1203.91^{+0.56}_{-0.56}$ & $8.6^{+2.7}_{-2.7}$ & $1334$ & $213^{+21}_{-21}$ (\textsuperscript{1}) & $20.02^{+1.45}_{-1.45}$ & $172.19^{+15.81}_{-15.81}$ & $5.37^{+0.65}_{-0.65}$ & $8.5$ & $5.3$ \\%
28 & $59060.54694786$ & $1201.91^{+1.90}_{-1.90}$ & $15.1^{+3.0}_{-3.0}$ & $1445$ & $187^{+19}_{-19}$ (\textsuperscript{1}) & $32.52^{+1.12}_{-1.12}$ & $490.07^{+21.50}_{-21.50}$ & $13.37^{+1.11}_{-1.11}$ & $13.8$ & $7.8$ \\%
29 & $59061.49706760$ & $1202.60^{+1.59}_{-1.59}$ & $5.5^{+2.6}_{-2.6}$ & $1280$ & $218^{+22}_{-22}$ (\textsuperscript{1}) & $22.40^{+1.00}_{-1.00}$ & $122.37^{+6.95}_{-6.95}$ & $3.89^{+0.41}_{-0.41}$ & $8.8$ & $5.5$ \\%
30 & $59061.50017397$ & $1203.81^{+1.57}_{-1.57}$ & $15.9^{+4.1}_{-4.1}$ & $1422$ & $57^{+6}_{-6}$ (\textsuperscript{1}) & $32.27^{+0.93}_{-0.93}$ & $513.57^{+18.87}_{-18.87}$ & $4.32^{+1.89}_{-1.89}$ & $11.1$ & $8.0$ \\%
31 & $59061.50327903$ & $1201.14^{+1.52}_{-1.52}$ & $11.4^{+2.6}_{-2.6}$ & $1388$ & $139^{+14}_{-14}$ (\textsuperscript{2}) & $23.02^{+0.74}_{-0.74}$ & $263.16^{+10.76}_{-10.76}$ & $5.35^{+0.51}_{-0.51}$ & $8.7$ & $5.8$ \\%
32 & $59061.50358025$ & $1201.70^{+1.84}_{-1.84}$ & $10.4^{+3.7}_{-3.7}$ & $1320$ & $271^{+27}_{-27}$ (\textsuperscript{1}) & $24.04^{+1.41}_{-1.41}$ & $250.91^{+18.73}_{-18.73}$ & $9.94^{+1.57}_{-1.57}$ & $7.3$ & $5.6$ \\%
33 & $59061.50535443$ & $1205.38^{+0.90}_{-0.90}$ & $9.4^{+3.0}_{-3.0}$ & $1428$ & $113^{+11}_{-11}$ (\textsuperscript{1}) & $25.15^{+0.79}_{-0.79}$ & $236.21^{+9.38}_{-9.38}$ & $3.92^{+0.41}_{-0.41}$ & $5.9$ & $3.4$ \\%
34 & $59061.50751035$ & $1201.17^{+1.79}_{-1.79}$ & $11.4^{+3.8}_{-3.8}$ & $1432$ & $86^{+9}_{-9}$ (\textsuperscript{1}) & $25.13^{+0.73}_{-0.73}$ & $286.56^{+10.62}_{-10.62}$ & $3.60^{+0.25}_{-0.25}$ & $8.4$ & $6.5$ \\%
35 & $59061.51067699$ & $1202.81^{+1.31}_{-1.31}$ & $13.3^{+3.7}_{-3.7}$ & $1432$ & $116^{+12}_{-12}$ (\textsuperscript{1}) & $27.01^{+0.82}_{-0.82}$ & $358.05^{+13.74}_{-13.74}$ & $6.09^{+1.45}_{-1.45}$ & $6.1$ & $4.2$ \\%
36 & $59061.51889186$ & $1202.36^{+1.85}_{-1.85}$ & $13.0^{+5.3}_{-5.3}$ & $1319$ & $130^{+13}_{-13}$ (\textsuperscript{2}) & $22.81^{+1.11}_{-1.11}$ & $297.36^{+18.39}_{-18.39}$ & $5.64^{+0.57}_{-0.57}$ & $6.2$ & $4.1$ \\%
37 & $59061.52526941$ & $1200.56^{+1.48}_{-1.48}$ & $9.6^{+2.3}_{-2.3}$ & $1470$ & $193^{+19}_{-19}$ (\textsuperscript{1}) & $23.75^{+0.77}_{-0.77}$ & $228.95^{+9.43}_{-9.43}$ & $6.47^{+0.97}_{-0.97}$ & $11.2$ & $7.8$ \\%
38 & $59061.53097863$ & $1201.52^{+1.56}_{-1.56}$ & $12.8^{+3.8}_{-3.8}$ & $1323$ & $147^{+15}_{-15}$ (\textsuperscript{2}) & $22.38^{+1.12}_{-1.12}$ & $285.42^{+18.11}_{-18.11}$ & $6.13^{+0.67}_{-0.67}$ & $10.3$ & $3.6$ \\%
39 & $59061.53278232$ & $1203.60^{+1.24}_{-1.24}$ & $7.1^{+3.1}_{-3.1}$ & $1426$ & $140^{+14}_{-14}$ (\textsuperscript{1}) & $19.47^{+0.63}_{-0.63}$ & $138.66^{+5.70}_{-5.70}$ & $2.83^{+0.32}_{-0.32}$ & $6.4$ & $4.5$ \\%
40 & $59061.54840199$ & $1203.37^{+4.02}_{-4.02}$ & $28.8^{+6.3}_{-6.3}$ & $1345$ & $305^{+30}_{-30}$ (\textsuperscript{2}) & $24.29^{+1.72}_{-1.72}$ & $700.01^{+62.87}_{-62.87}$ & $31.17^{+3.46}_{-3.46}$ & $8.1$ & $6.5$ \\%
41 & $59067.46362248$ & $1202.50^{+1.98}_{-1.98}$ & $12.2^{+4.2}_{-4.2}$ & $1442$ & $92^{+9}_{-9}$ (\textsuperscript{2}) & $25.70^{+0.77}_{-0.77}$ & $313.39^{+11.97}_{-11.97}$ & $4.20^{+0.69}_{-0.69}$ & $7.2$ & $5.7$ \\%
42 & $59067.47196287$ & $1201.75^{+1.87}_{-1.87}$ & $8.6^{+3.2}_{-3.2}$ & $1436$ & $47^{+5}_{-5}$ (\textsuperscript{1}) & $32.37^{+0.81}_{-0.81}$ & $277.86^{+8.88}_{-8.88}$ & $1.90^{+0.21}_{-0.21}$ & $6.4$ & $4.2$ \\%
43 & $59067.48018037$ & $1203.08^{+1.81}_{-1.81}$ & $3.5^{+2.5}_{-2.5}$ & $1255$ & $150^{+15}_{-15}$ (\textsuperscript{1}) & $29.22^{+1.62}_{-1.62}$ & $101.83^{+7.17}_{-7.17}$ & $2.24^{+0.31}_{-0.31}$ & $10.5$ & $8.1$ \\%
44 & $59067.48114672$ & $1203.99^{+0.69}_{-0.69}$ & $5.7^{+1.3}_{-1.3}$ & $1427$ & $209^{+53}_{-53}$ (\textsuperscript{1}) & $30.03^{+1.05}_{-1.05}$ & $170.94^{+7.56}_{-7.56}$ & $5.21^{+0.63}_{-0.63}$ & $16.0$ & $10.8$ \\%
45 & $59067.48472680$ & $1201.74^{+1.98}_{-1.98}$ & $7.3^{+2.8}_{-2.8}$ & $1284$ & $125^{+13}_{-13}$ (\textsuperscript{1}) & $43.75^{+2.23}_{-2.23}$ & $318.49^{+20.58}_{-20.58}$ & $5.84^{+0.76}_{-0.76}$ & $7.5$ & $6.3$ \\%
46 & $59067.48919208$ & $1199.76^{+1.58}_{-1.58}$ & $8.5^{+4.1}_{-4.1}$ & $1378$ & $233^{+23}_{-23}$ (\textsuperscript{1}) & $23.11^{+1.71}_{-1.71}$ & $195.47^{+18.32}_{-18.32}$ & $6.66^{+0.93}_{-0.93}$ & $6.8$ & $4.7$ \\%
47 & $59067.48943302$ & $1201.86^{+1.99}_{-1.99}$ & $5.6^{+0.8}_{-0.8}$ & $1391$ & $158^{+7}_{-7}$ (\textsuperscript{2}) & $49.60^{+1.64}_{-1.64}$ & $277.68^{+11.69}_{-11.69}$ & $6.43^{+0.39}_{-0.39}$ & $17.2$ & $11.9$ \\%
48 & $59067.49143691$ & $1202.80^{+1.89}_{-1.89}$ & $11.9^{+3.4}_{-3.4}$ & $1390$ & $181^{+18}_{-18}$ (\textsuperscript{2}) & $18.55^{+0.59}_{-0.59}$ & $221.24^{+8.98}_{-8.98}$ & $5.84^{+1.07}_{-1.07}$ & $8.1$ & $5.9$ \\%
49 & $59067.50320604$ & $1201.08^{+1.46}_{-1.46}$ & $17.9^{+2.6}_{-2.6}$ & $1382$ & $146^{+15}_{-15}$ (\textsuperscript{2}) & $35.60^{+1.14}_{-1.14}$ & $637.65^{+25.94}_{-25.94}$ & $13.65^{+0.68}_{-0.68}$ & $11.9$ & $8.6$ \\%
50 & $59067.50475553$ & $1202.20^{+1.89}_{-1.89}$ & $9.4^{+2.4}_{-2.4}$ & $1395$ & $243^{+24}_{-24}$ (\textsuperscript{1}) & $28.44^{+1.62}_{-1.62}$ & $268.47^{+19.43}_{-19.43}$ & $9.54^{+1.23}_{-1.23}$ & $11.5$ & $7.0$ \\%
51 & $59067.50694658$ & $1204.42^{+1.32}_{-1.32}$ & $13.2^{+1.9}_{-1.9}$ & $1400$ & $129^{+13}_{-13}$ (\textsuperscript{2}) & $40.97^{+1.44}_{-1.44}$ & $538.88^{+24.06}_{-24.06}$ & $10.19^{+1.69}_{-1.69}$ & $13.6$ & $9.0$ \\%
52 & $59067.51160959$ & $1201.35^{+1.71}_{-1.71}$ & $8.6^{+2.7}_{-2.7}$ & $1404$ & $161^{+16}_{-16}$ (\textsuperscript{2}) & $21.48^{+0.74}_{-0.74}$ & $184.78^{+8.04}_{-8.04}$ & $4.35^{+0.68}_{-0.68}$ & $7.8$ & $4.0$ \\%
53 & $59067.51809977$ & $1201.94^{+1.90}_{-1.90}$ & $5.9^{+1.8}_{-1.8}$ & $1424$ & $97^{+10}_{-10}$ (\textsuperscript{1}) & $29.65^{+0.92}_{-0.92}$ & $175.87^{+6.91}_{-6.91}$ & $2.49^{+0.40}_{-0.40}$ & $7.9$ & $5.6$ \\%
54 & $59067.51855025$ & $1201.72^{+1.91}_{-1.91}$ & $14.1^{+3.9}_{-3.9}$ & $1270$ & $123^{+12}_{-12}$ (\textsuperscript{1}) & $28.56^{+1.53}_{-1.53}$ & $402.14^{+27.39}_{-27.39}$ & $7.24^{+0.62}_{-0.62}$ & $6.1$ & $4.2$ \\%
55 & $59067.51888109$ & $1201.47^{+1.10}_{-1.10}$ & $9.2^{+2.3}_{-2.3}$ & $1193$ & $353^{+35}_{-35}$ (\textsuperscript{1}) & $19.43^{+0.88}_{-0.88}$ & $178.33^{+10.28}_{-10.28}$ & $9.22^{+0.71}_{-0.71}$ & $8.1$ & $6.9$ \\%
56 & $59067.52572300$ & $1204.86^{+0.97}_{-0.97}$ & $6.6^{+1.3}_{-1.3}$ & $1390$ & $365^{+67}_{-67}$ (\textsuperscript{1}) & $30.08^{+2.02}_{-2.02}$ & $197.50^{+16.87}_{-16.87}$ & $10.53^{+1.22}_{-1.22}$ & $17.0$ & $12.1$ \\%
57 & $59067.52790177$ & $1198.76^{+0.75}_{-0.75}$ & $15.6^{+4.5}_{-4.5}$ & $1287$ & $232^{+23}_{-23}$ (\textsuperscript{1}) & $23.17^{+1.36}_{-1.36}$ & $361.10^{+26.82}_{-26.82}$ & $12.25^{+1.07}_{-1.07}$ & $8.4$ & $5.9$ \\%
58 & $59067.52795093$ & $1203.76^{+4.44}_{-4.44}$ & $22.0^{+4.4}_{-4.4}$ & $1199$ & $394^{+39}_{-39}$ (\textsuperscript{1}) & $22.88^{+1.17}_{-1.17}$ & $503.46^{+32.72}_{-32.72}$ & $28.98^{+2.80}_{-2.80}$ & $11.4$ & $9.1$ \\%
59 & $59067.52812338$ & $1201.84^{+2.07}_{-2.07}$ & $9.2^{+2.3}_{-2.3}$ & $1327$ & $352^{+35}_{-35}$ (\textsuperscript{1}) & $21.40^{+1.44}_{-1.44}$ & $196.93^{+16.85}_{-16.85}$ & $10.12^{+1.02}_{-1.02}$ & $11.9$ & $8.9$ \\%
60 & $59067.52833961$ & $1201.59^{+4.76}_{-4.76}$ & $28.6^{+7.5}_{-7.5}$ & $1376$ & $127^{+13}_{-13}$ (\textsuperscript{2}) & $24.16^{+0.83}_{-0.83}$ & $689.75^{+30.19}_{-30.19}$ & $12.83^{+1.62}_{-1.62}$ & $5.1$ & $4.5$ \\%
61 & $59067.53192495$ & $1202.55^{+1.92}_{-1.92}$ & $10.8^{+1.6}_{-1.6}$ & $1338$ & $392^{+38}_{-38}$ (\textsuperscript{1}) & $33.11^{+2.18}_{-2.18}$ & $358.15^{+30.01}_{-30.01}$ & $20.52^{+1.91}_{-1.91}$ & $16.6$ & $12.6$ \\%
62 & $59077.43170549$ & $1200.04^{+1.14}_{-1.14}$ & $8.9^{+2.9}_{-2.9}$ & $1249$ & $135^{+14}_{-14}$ (\textsuperscript{2}) & $26.92^{+1.06}_{-1.06}$ & $240.59^{+11.98}_{-11.98}$ & $4.75^{+1.47}_{-1.47}$ & $9.4$ & $8.9$ \\%
63 & $59077.46064368$ & $1201.53^{+1.91}_{-1.91}$ & $6.9^{+3.5}_{-3.5}$ & $1396$ & $257^{+26}_{-26}$ (\textsuperscript{1}) & $26.12^{+1.39}_{-1.39}$ & $179.23^{+12.10}_{-12.10}$ & $6.74^{+0.67}_{-0.67}$ & $9.2$ & $7.0$ \\%
64 & $59077.46596665$ & $1200.95^{+1.63}_{-1.63}$ & $11.6^{+4.4}_{-4.4}$ & $1075$ & $230^{+23}_{-23}$ (\textsuperscript{1}) & $22.50^{+0.65}_{-0.65}$ & $261.37^{+9.66}_{-9.66}$ & $8.77^{+0.72}_{-0.72}$ & $7.2$ & $4.8$ \\%
65 & $59077.46688247$ & $1203.02^{+1.60}_{-1.60}$ & $8.2^{+3.0}_{-3.0}$ & $1320$ & $380^{+38}_{-38}$ (\textsuperscript{1}) & $15.28^{+0.79}_{-0.79}$ & $125.19^{+8.20}_{-8.20}$ & $6.96^{+0.55}_{-0.55}$ & $10.7$ & $7.1$ \\%
66 & $59077.47341161$ & $1202.81^{+1.80}_{-1.80}$ & $10.2^{+2.4}_{-2.4}$ & $1414$ & $170^{+17}_{-17}$ (\textsuperscript{2}) & $24.10^{+0.80}_{-0.80}$ & $246.49^{+10.40}_{-10.40}$ & $6.13^{+0.36}_{-0.36}$ & $9.0$ & $4.4$ \\%
67 & $59077.47762721$ & $1199.90^{+1.39}_{-1.39}$ & $6.2^{+1.8}_{-1.8}$ & $1428$ & $137^{+14}_{-14}$ (\textsuperscript{2}) & $30.67^{+1.01}_{-1.01}$ & $188.83^{+7.90}_{-7.90}$ & $3.77^{+0.52}_{-0.52}$ & $10.6$ & $7.4$ \\%
68 & $59077.48005306$ & $1200.99^{+2.25}_{-2.25}$ & $13.6^{+3.3}_{-3.3}$ & $1138$ & $194^{+19}_{-19}$ (\textsuperscript{2}) & $27.20^{+0.72}_{-0.72}$ & $371.14^{+12.56}_{-12.56}$ & $10.55^{+0.95}_{-0.95}$ & $6.7$ & $5.6$ \\%
69 & $59077.48829166$ & $1202.37^{+1.90}_{-1.90}$ & $9.7^{+2.9}_{-2.9}$ & $1290$ & $196^{+20}_{-20}$ (\textsuperscript{2}) & $22.00^{+1.02}_{-1.02}$ & $213.82^{+12.62}_{-12.62}$ & $6.12^{+0.44}_{-0.44}$ & $6.6$ & $4.5$ \\%
70 & $59089.42255268$ & $1202.71^{+4.08}_{-4.08}$ & $31.3^{+2.6}_{-2.6}$ & $1402$ & $121^{+12}_{-12}$ (\textsuperscript{2}) & $37.39^{+1.44}_{-1.44}$ & $1170.41^{+14.29}_{-14.29}$ & $20.64^{+1.99}_{-1.99}$ & $18.4$ & $11.7$ \\%
71 & $59089.42638332$ & $1202.47^{+3.64}_{-3.64}$ & $24.6^{+2.3}_{-2.3}$ & $1396$ & $214^{+21}_{-21}$ (\textsuperscript{1}) & $25.25^{+0.98}_{-0.98}$ & $620.72^{+7.69}_{-7.69}$ & $19.44^{+2.30}_{-2.30}$ & $19.0$ & $8.1$ \\%
72 & $59089.44362517$ & $1202.58^{+4.56}_{-4.56}$ & $17.5^{+3.8}_{-3.8}$ & $1399$ & $106^{+11}_{-11}$ (\textsuperscript{2}) & $18.50^{+0.61}_{-0.61}$ & $324.46^{+3.38}_{-3.38}$ & $5.04^{+1.33}_{-1.33}$ & $11.3$ & $8.3$ \\%
73 & $59089.44381959$ & $1202.25^{+4.69}_{-4.69}$ & $19.5^{+2.1}_{-2.1}$ & $1388$ & $154^{+6}_{-6}$ (\textsuperscript{2}) & $31.72^{+1.04}_{-1.04}$ & $617.81^{+6.43}_{-6.43}$ & $13.91^{+0.52}_{-0.52}$ & $16.8$ & $11.8$ \\%
74 & $59089.45120231$ & $1205.94^{+2.67}_{-2.67}$ & $10.3^{+4.1}_{-4.1}$ & $1269$ & $341^{+34}_{-34}$ (\textsuperscript{1}) & $10.25^{+0.65}_{-0.65}$ & $105.19^{+2.11}_{-2.11}$ & $5.25^{+0.41}_{-0.41}$ & $11.1$ & $9.3$ \\%
75 & $59111.36402893$ & $1202.22^{+4.42}_{-4.42}$ & $11.9^{+3.7}_{-3.7}$ & $1389$ & $221^{+22}_{-22}$ (\textsuperscript{2}) & $12.76^{+0.54}_{-0.54}$ & $151.20^{+4.10}_{-4.10}$ & $4.88^{+4.32}_{-4.32}$ & $8.3$ & $5.2$ \\%
76 & $59111.36526831$ & $1201.45^{+1.72}_{-1.72}$ & $17.3^{+3.6}_{-3.6}$ & $1398$ & $151^{+15}_{-15}$ (\textsuperscript{1}) & $16.49^{+0.52}_{-0.52}$ & $284.48^{+5.74}_{-5.74}$ & $6.29^{+3.40}_{-3.40}$ & $9.8$ & $7.6$ \\%
77 & $59111.36856187$ & $1203.11^{+1.63}_{-1.63}$ & $14.7^{+1.0}_{-1.0}$ & $1393$ & $220^{+19}_{-19}$ (\textsuperscript{1}) & $55.27^{+2.29}_{-2.29}$ & $811.03^{+21.41}_{-21.41}$ & $26.03^{+1.22}_{-1.22}$ & $29.1$ & $19.5$ \\%
78 & $59111.37352919$ & $1202.87^{+1.71}_{-1.71}$ & $9.4^{+2.2}_{-2.2}$ & $1403$ & $216^{+22}_{-22}$ (\textsuperscript{1}) & $18.43^{+0.76}_{-0.76}$ & $173.94^{+4.55}_{-4.55}$ & $5.50^{+0.57}_{-0.57}$ & $7.7$ & $6.0$ \\%
79 & $59111.37932233$ & $1200.55^{+1.64}_{-1.64}$ & $11.2^{+1.7}_{-1.7}$ & $1375$ & $187^{+19}_{-19}$ (\textsuperscript{2}) & $23.60^{+0.99}_{-0.99}$ & $264.41^{+7.08}_{-7.08}$ & $7.23^{+0.66}_{-0.66}$ & $12.5$ & $8.8$ \\%
\end{longtable}
\end{center}
\begin{tablenotes}
\item[a)] $a)$ The time of arrival of the pulses, corrected to an infinite frequency. \\%
\item[b)] $b)$ The structure-optimized DM measured by using the \texttt{DM-Power} algorithm \citep{lin2022dm} (available at https://github.com/hsiuhsil/DM-power). \\%
\item[c)] $c)$ The equivalent width, W$_{eq}$, is defined as the width of a rectangular pulse that has an area equal to the integral of the observed pulse and a height matching the peak flux density.
\\%
\item[d)] $d)$ Two methods are employed to measure the bandwidth: $^{\#}$1) Fitting a Gaussian function to the pulse's bandpass to determine the bandwidth, with the 1$\sigma$ measurement uncertainty. $^{\#}$2) Calculating the cumulative distribution function (CDF) of the bandpass and determining the bandwidth based on the derivative of the CDF with respect to the frequency, with the 1$\sigma$ measurement uncertainty obtained through bootstrapping. For pulses with \( SNR_\text{A} \leq 10 \), a conservative estimate of uncertainty of 20\% is assumed. \\%
\item[e)] $e)$ The equivalent isotropic energies of the pulses are calculated as described in equation (9) of \citet{zhang2018fast} (see Eq.~(\ref{eq:energy})). \\%
\item[f)] $f)$ The $SNR_\text{A}$ and $SNR_\text{B}$ are obtained using \texttt{single\_pulse\_search.py} from \texttt{PRESTO}, which searches the two time series created separately by the BASSET and the standard pipeline. \\%
\end{tablenotes}

\bibliography{paper}{}
\bibliographystyle{aasjournal}



\end{document}